\documentclass[aps,prd,reprint,amsmath,amssymb,floats,floatfix, twocolumn,
superscriptaddress,nofootinbib,showpacs]{revtex4-1}

\pdfoutput=1

\usepackage[usenames,dvipsnames]{xcolor}
\usepackage{hyperref}
\definecolor{cobalt}{RGB}{44, 98, 120}
\hypersetup{
  colorlinks,
  citecolor=Violet,
  linkcolor=cobalt,
  urlcolor=Blue
}
\usepackage[capitalise]{cleveref}
\usepackage[normalem]{ulem}
\usepackage[caption=false]{subfig}
\usepackage{tensor}
\usepackage{soul}
\usepackage{microtype}

\usepackage{tikz}
\usetikzlibrary{positioning}
\usetikzlibrary{decorations.pathmorphing}
\usetikzlibrary{decorations.text}
\usetikzlibrary{shapes.misc}
\usetikzlibrary{calc}
\usepackage{tikzscale}

\graphicspath{{figures/}}

\newcommand{\unit}[1]{\ensuremath{\, \mathrm{#1}}}

\newcommand{\Cornell}{\affiliation{Cornell Center for Astrophysics
    and Planetary Science, Cornell University, Ithaca, New York 14853, USA}}

\newcommand{\ts}{\textsuperscript}
\newcommand{\SpEC}{\textsc{SpEC}}

\begin{document}

\title{
  A Parallel Adaptive Event Horizon Finder for Numerical Relativity
}

\author{Andy Bohn}
\email[Contact email: ]{adb228@cornell.edu}
\author{Lawrence E. Kidder, Saul A. Teukolsky} \Cornell

\date{\today}

\begin{abstract}
With Advanced LIGO detecting the gravitational waves emitted from a pair of
merging black holes in late 2015, we have a new perspective into the
strong field regime of binary black hole systems.
Event horizons are the defining features of such black hole spacetimes.
We introduce a new code for locating event horizons in numerical simulations
based on a Delaunay triangulation on a topological sphere.
The code can automatically refine arbitrary regions of the
event horizon surface
to find and explore features such as the hole in a toroidal event horizon,
as discussed in
our
companion paper.
We also investigate various ways of integrating the geodesic equation and find
evolution equations that can be integrated efficiently with high accuracy.

\end{abstract}


\maketitle

\newcommand{\tikzprefixA}{Tikz}
\section{Introduction}

In late 2015, the Advanced LIGO interferometers detected
the gravitational radiation from a pair of merging black
holes~\cite{Abbott:2016blz}.
This observation gives a unique view into the highly nonlinear
regime of compact-object binary mergers, and the observed gravitational
waveform is
entirely consistent with General Relativity~\cite{TheLIGOScientific:2016src}.
While numerical relativity simulations help with detecting and analyzing
signals that Advanced LIGO receives, they also provide a laboratory for
exploring
the entire compact object coalescence parameter space, including
the $7$-dimensional space of binary black hole (BBH) mergers.
Algorithmic improvements in addition to increasing computational power
over time have led to a large surge in the number of BBH simulations
available to the community~\cite{
Pekowsky:2013ska, SXSCatalog, Ajith:2012az, Zlochower2015, Bruegmann2006,
Jani:2016wkt}.

Among the properties of the spacetime that can be studied
using numerical simulations, perhaps the most interesting are those
of black hole
\textit{event horizons} (EH), the
boundaries of the causal past of future null infinity.
The EH surface is therefore dependent on the entire future of the spacetime,
making it impossible to locate during BBH simulations that progress forwards
in time.
A similar surface, called the
\textit{apparent horizon} (AH), is the boundary between outward directed light
rays
moving away from or toward the center of the black hole.
In particular, the EH always contains the AH if it exists, and the surfaces
are equal if the black hole has settled down to equilibrium.
Locating an AH at a certain time requires only information at that time,
so AHs are commonly located during BBH simulations as an EH substitute.
Even though EHs are more difficult to locate, we
are interested in how to find them because they define the surface of black
holes, and physical properties
such as the mass and angular momentum of black holes are determined
by integrations over the event horizon surface~\cite{PoissonToolkit}.

We locate event horizons in BBH mergers by utilizing a theorem
that the event horizon is generated by null geodesics having no
future end point~\cite{Penrose1968, hawkingellis, Wald84}.
Long after the black holes have merged, the spacetime settles down
to Kerr, where the EH is identical to the AH.
So we can select a set of outgoing null
geodesics that
lie on the apparent horizon of the remnant black hole near the end of the
BBH simulation~\cite{Anninos1995} and
integrate the geodesics backwards through time~\cite{Shapiro-Teukolsky:1980,
  Hughes1994,
  Shapiro1995, Anninos1995, Libson96, CohenPfeiffer2008, Cohen2012}.
The convention that we will follow in this paper
is to call these geodesics event horizon generators,
though they are only very good approximations to the true
generators~\cite{Cohen2012}.
Although generators of the horizon have no future endpoint, while tracing
the generators backwards in time, some may ``leave'' the event horizon
surface where they meet other generators of the horizon.
These points
are called \textit{caustics} when infinitesimally neighboring generators
join together, and \textit{crossover points} when non-neighboring
generators cross
paths~\cite{Shapiro1995, Siino1998b, Husa-Winicour:1999, Lehner1999, Cohen2012}.
After they leave the event horizon surface backwards in time, generators
are known as \textit{future generators} of the horizon.
When viewing the event horizon forwards in time, future generators become
generators of the event horizon after they join through either caustics or
crossover points.

The previous generation of event horizon finding code in \SpEC{}~\cite{
CohenPfeiffer2008, Cohen2012} was sufficient to locate event horizons
reasonably
accurately, but lacked the ability to adaptively refine itself to study
small scale features of the EH surface.
An example of a small scale feature we are interested in exploring
is a topological hole through the event horizon surface, causing the
EH topology to be toroidal.
The companion to this paper~\cite{BohnTorus2016} focuses on locating such
short-lived toroidal event horizons.
This paper outlines the details behind our new
event horizon finder, and the adaptive refinement tools that
are essential to resolve a toroidal event horizon.

The organization of this paper is as follows: In \cref{secOverview}
we give an overview of the backwards geodesic method for locating
event horizons.
In \cref{secTriangulation} we present the Delaunay triangulation~\cite{Delaunay,
Press2007}
on a spherical topology that we use to represent the event horizon surface,
allowing for adaptive refinement.
In \cref{secEvolution} and \cref{secInterpolation}, we show efficient null
geodesic evolution equations and outline how
we handle metric data during generator evolution.
In \cref{secInitialData}, we describe the initial data calculation
for event horizon generators, and in \cref{secIdentifyingGenerators} we
describe how we identify future
generators during the backwards in time evolution.

\section{Backwards geodesic method overview}
\label{secOverview}

Cohen~\textit{et al.}~\cite{CohenPfeiffer2008} compared three methods
for locating event horizons and found the most robust method to be the
backwards geodesic method.
We follow this approach,
where we evolve a set of event horizon
generators backwards in time to trace out the EH surface.
The generators are outward null geodesics that exponentially converge to
the true EH surface when traced backwards through time.
As we will discuss in \cref{secTriangulation},
we connect the generators together to form a polygon approximating
a smooth surface with the topology
of a sphere that may be self-intersecting.
This surface does not approximate the event horizon only, but
represents the union of the true event horizon and
the locus of future generators~\cite{ThorneSuggestion}.

\begin{figure}
  \centering
  \begin{tabular}{ccc}
    \includegraphics[width=0.33\columnwidth]{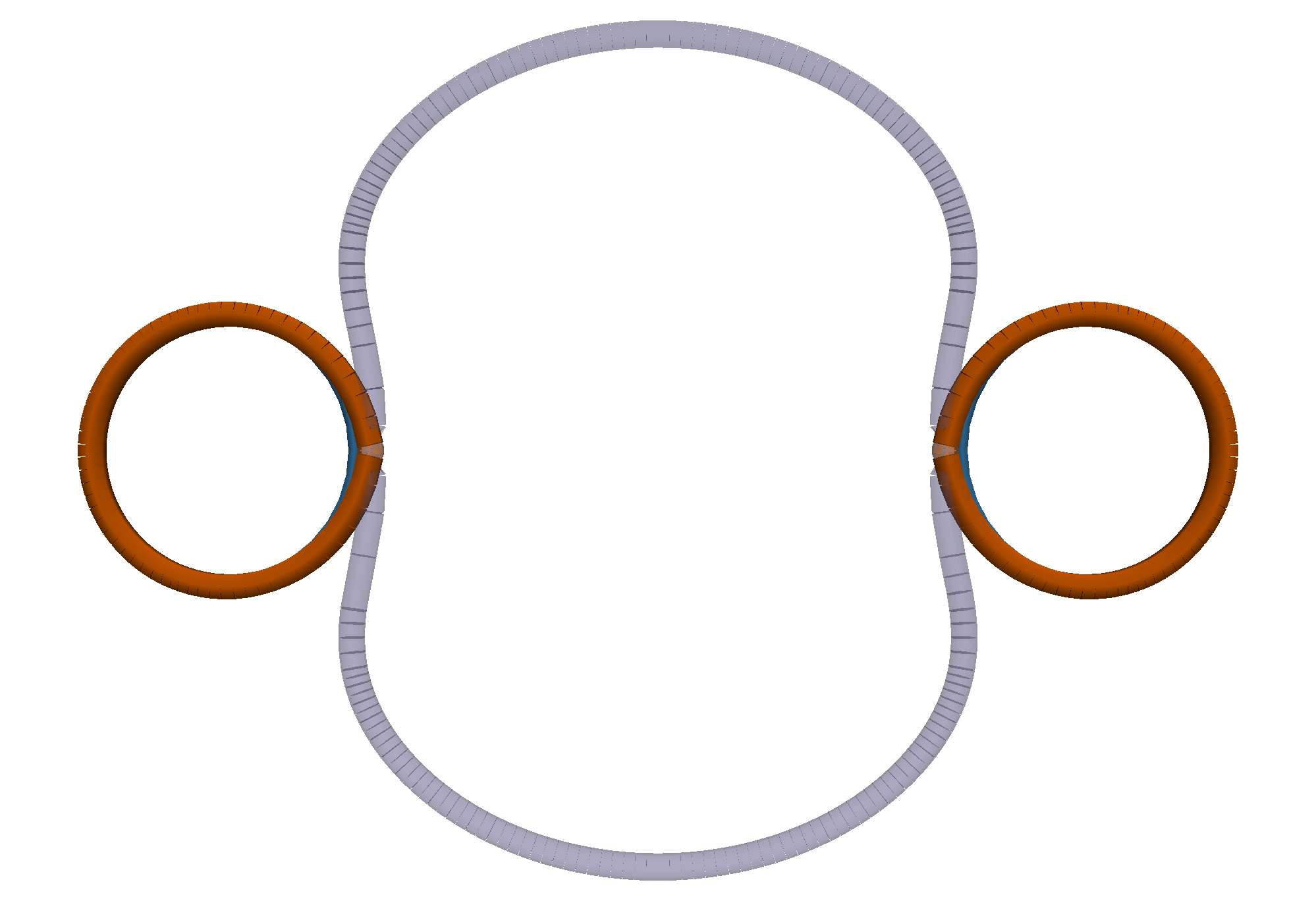} &
    \includegraphics[width=0.33\columnwidth]{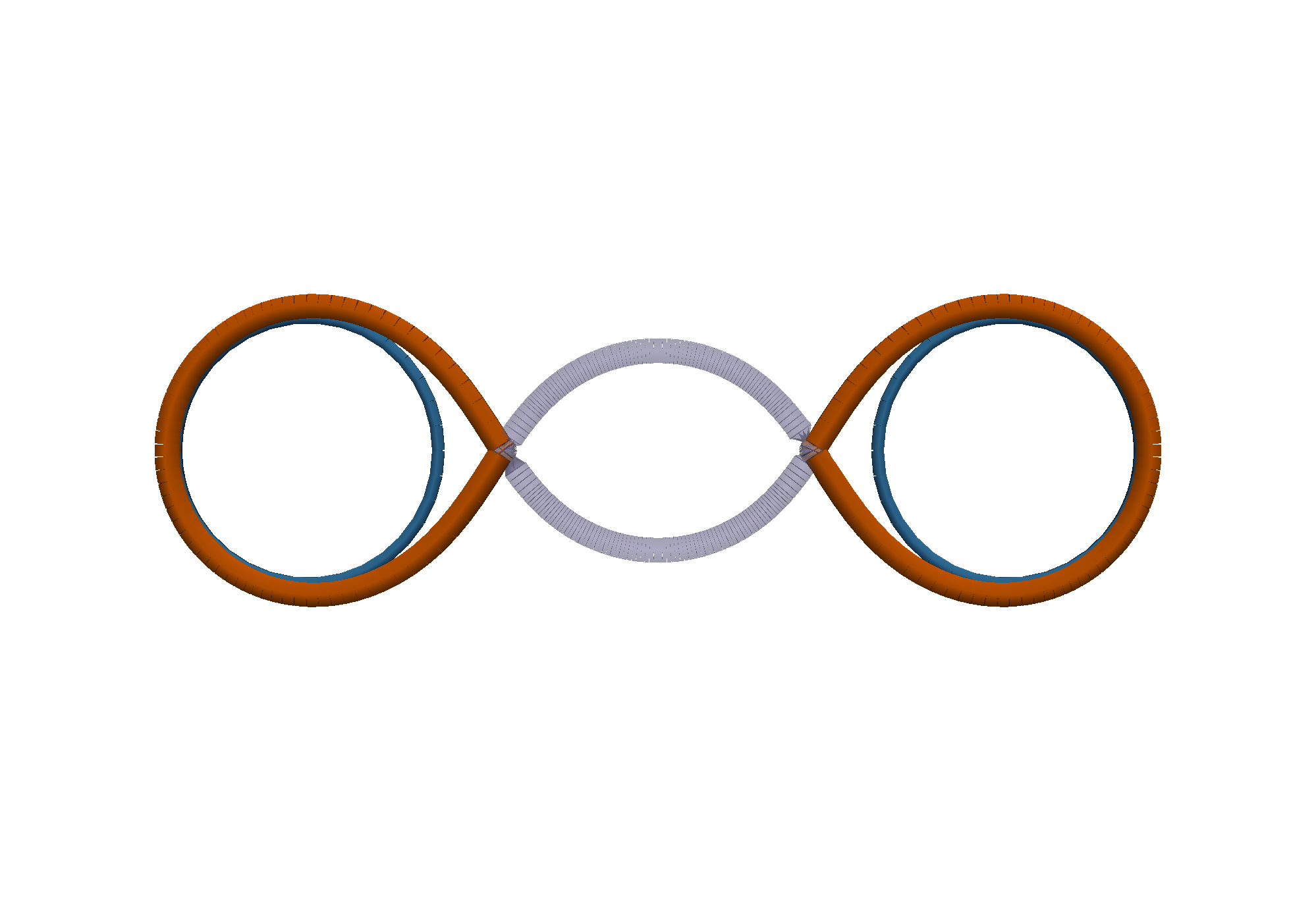} &
    \includegraphics[width=0.33\columnwidth]{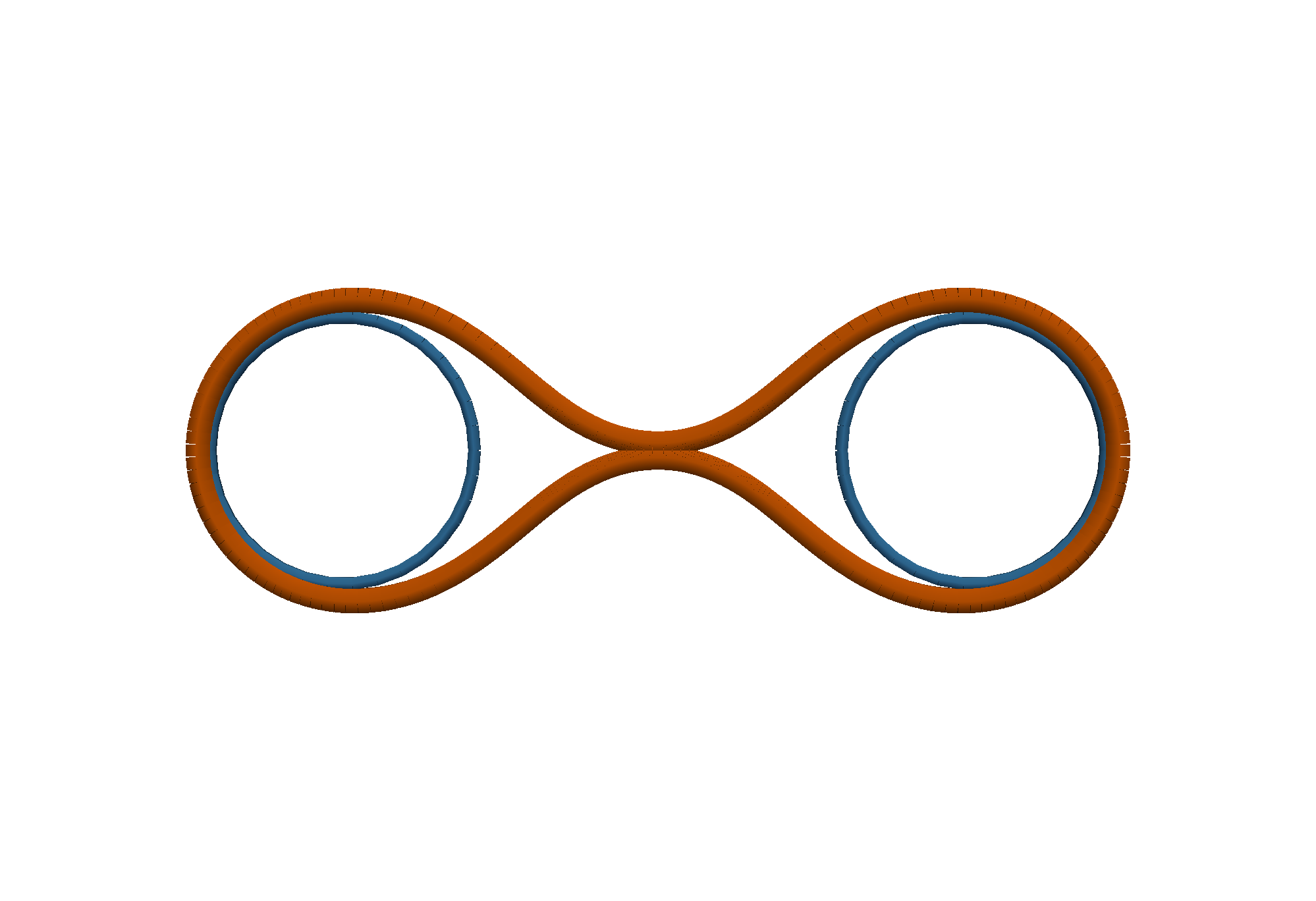} \\
    (a) $t=414.000 M$ &
    (b) $t=416.500 M$ &
    (c) $t=417.500 M$ \\
    \includegraphics[width=0.33\columnwidth]{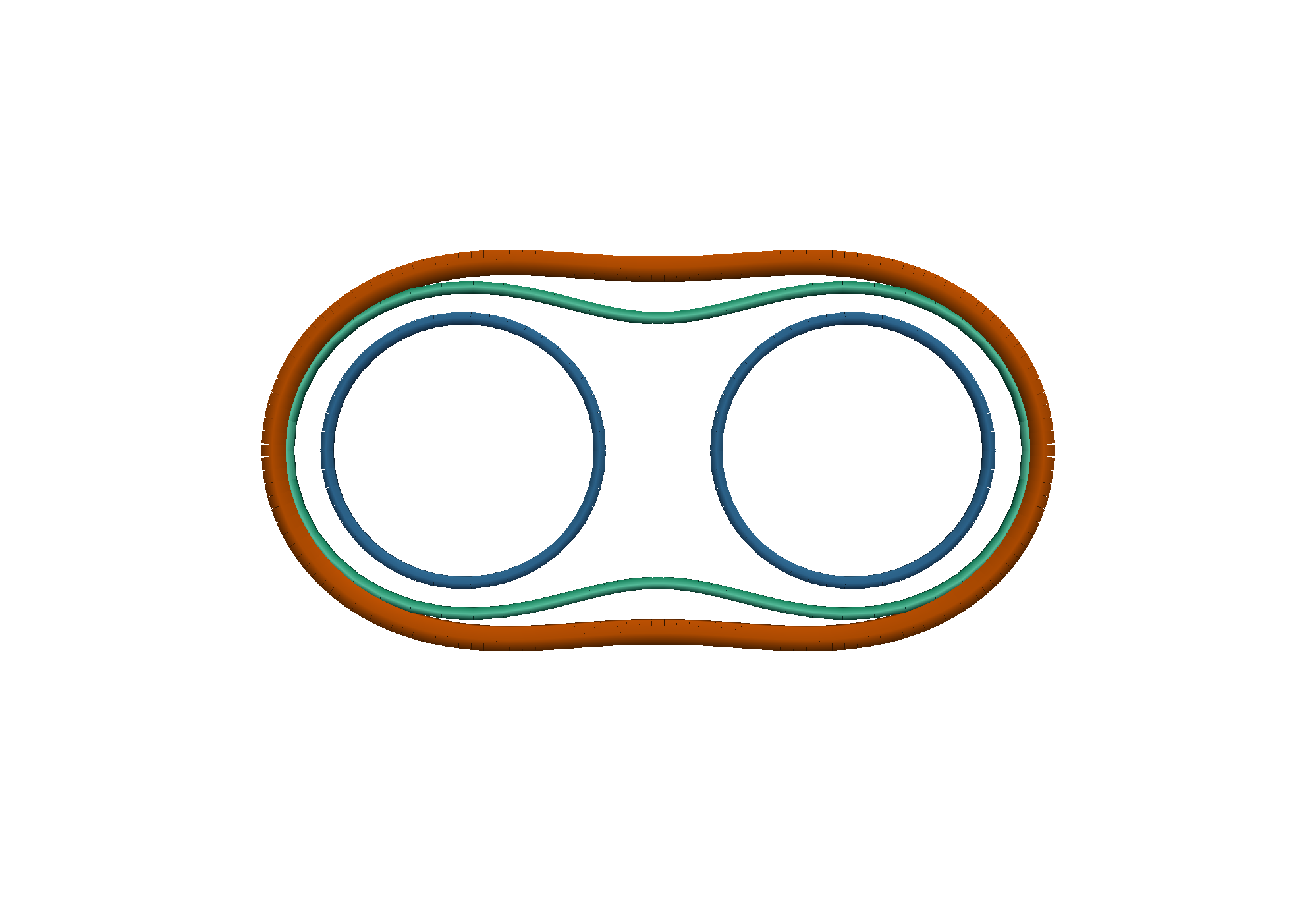} &
    \includegraphics[width=0.33\columnwidth]{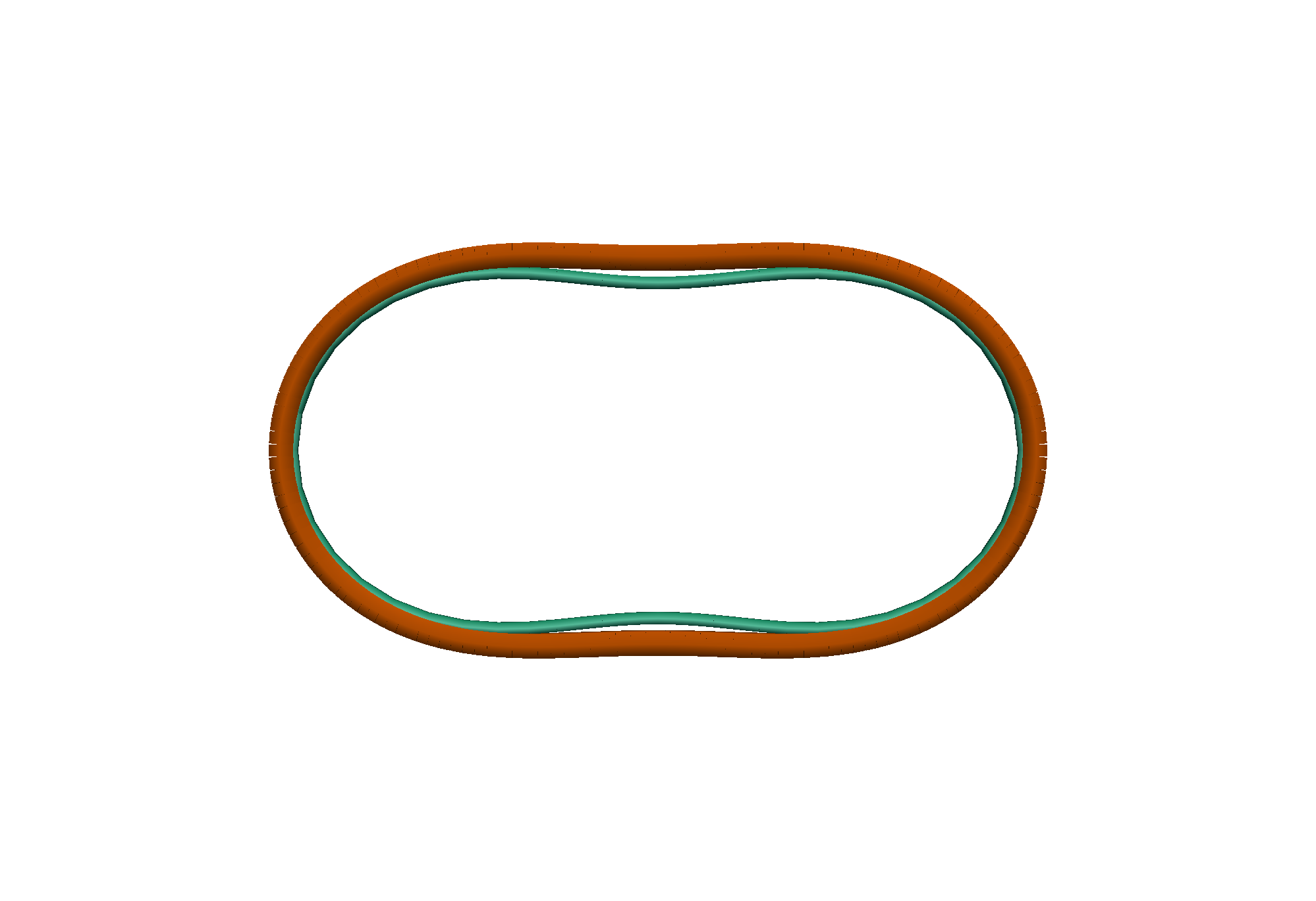} &
    \includegraphics[width=0.33\columnwidth]{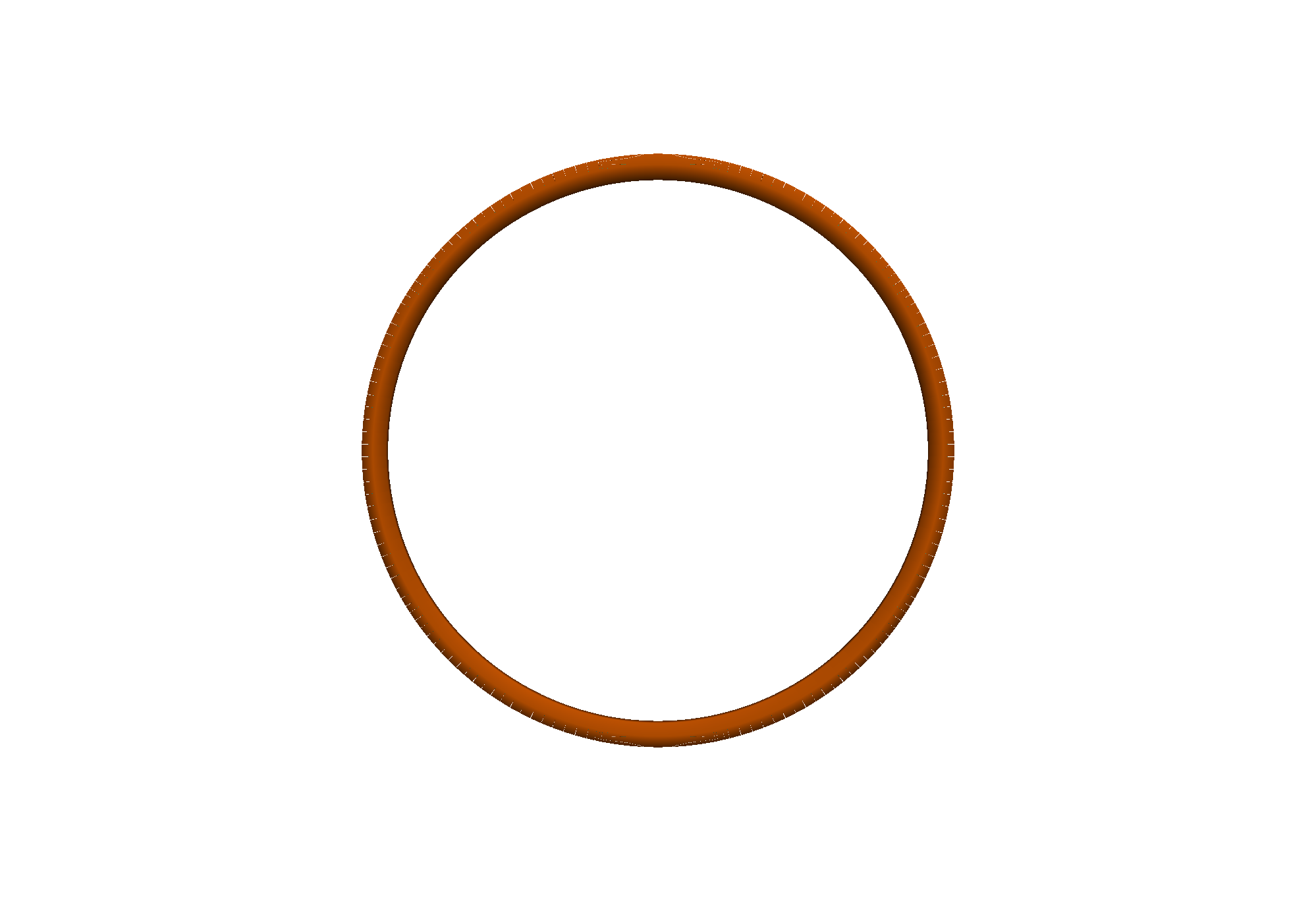} \\
    (d) $t=420.003 M$ &
    (e) $t=420.266 M$ &
    (f) $t=470.639 M$
  \end{tabular}
  \caption[Cross-sections of horizons for a head-on BBH merger]{
    Cross-sections through apparent horizons and the locus of event horizon
    generators
    for a head-on BBH merger, similar to Fig.~1 of~\cite{CohenPfeiffer2008}.
    Shown in translucent purple are future generators of the horizon that
    continuously merge onto the event horizon, shown in orange, until the
    merger in panel~(c).
    Shown as blue curves in panels~(a-d) are apparent horizons associated
    with the two individual black holes, and shown as a green curve in
    panels~(d-f) is a common apparent horizon.
  }
  \label{figThorneSuggestion}
\end{figure}

To make the discussion concrete, consider a head-on equal mass binary black
hole merger, shown in \cref{figThorneSuggestion}.
We see spatial cross-sections of
apparent horizon surfaces shown blue or green, event horizon surfaces
shown in orange, and the future
generator surface shown in translucent purple.
In panel~(a), sufficiently long before the merger, the event horizon
surfaces are almost identical to the blue apparent horizon surfaces,
which are hardly visible at this time.
The future generator surface consists of future generators that will
join onto the event horizon surface in the future.
When rotating this panel about the rotational axis of symmetry, the union
of the event horizon surfaces and future generator surface forms a smooth
topological sphere.
In panel~(b), shortly before the merger, the future generator surface
is smaller because some of the future generators joined the event horizon
between this panel and the previous panel.
We can see the difference between the AH and EH
surfaces increases as we get closer to the merger.
There are no more future generators in panel~(c) since they have all
joined the event horizon surface.

In panel~(d), a common apparent horizon shown in green has formed around the
two interior apparent horizons, and all three apparent horizons lie entirely
on or within the event horizon, as they should.
As time progresses to panels~(e)~and~(f), we stop tracking the blue inner
apparent horizons, the EH settles to a stationary state,
and the common AH approaches the event horizon until
the two surfaces eventually coincide.
With this picture in mind, the method used to locate the EH
is to evolve generators backwards in time
from panel~(f) toward panel~(a), which traces out the union of the
event horizon surface with the future generator surface.
Backwards in time, some generators ``leave'' the
event horizon surface as seen in panels~(b)~and~(a), so we must
be able to identify which generators leave the surface and when they leave.

\section{Event horizon representation}
\label{secTriangulation}

One of the shortcomings of our previous event horizon finder was the lack of
flexibility in refining the distribution of event horizon
generators in certain regions of interest.
The method of distributing event horizon generators in
Cohen~\textit{et al.}~\cite{CohenPfeiffer2008}
used collocation
points in a spherical harmonic ($Y_{lm}(u, v)$) expansion,
with $u$ values chosen so that $\cos{u}$ were the roots of the Legendre
polynomial of order $L+1$, and $v$ values uniformly distributed in
$[0, 2\pi)$, yielding $2(L+1)^2$ generators.
This results in the generators not being distributed evenly over the event
horizon surface, and does not allow one
to increase the resolution of a small patch of
the surface.

We want to be able to evenly distribute event horizon generators
over the event horizon as well as to be able to adaptively refine
regions of the surface to sufficiently resolve the small scale features
of the merger.
Compared to other methods of locating event horizons~\cite{CohenPfeiffer2008},
the backwards geodesic method
allows simple adaptive refinement, in that
we only need to add more generators wherever we want to refine.
In addition to being able to place generators where desired, we require
of our EH representation
the ability to connect the generators to approximate a smooth surface.

\begin{figure*}
  \centering
  \begin{tabular}{ccc}
    \includegraphics[width=0.33\textwidth]{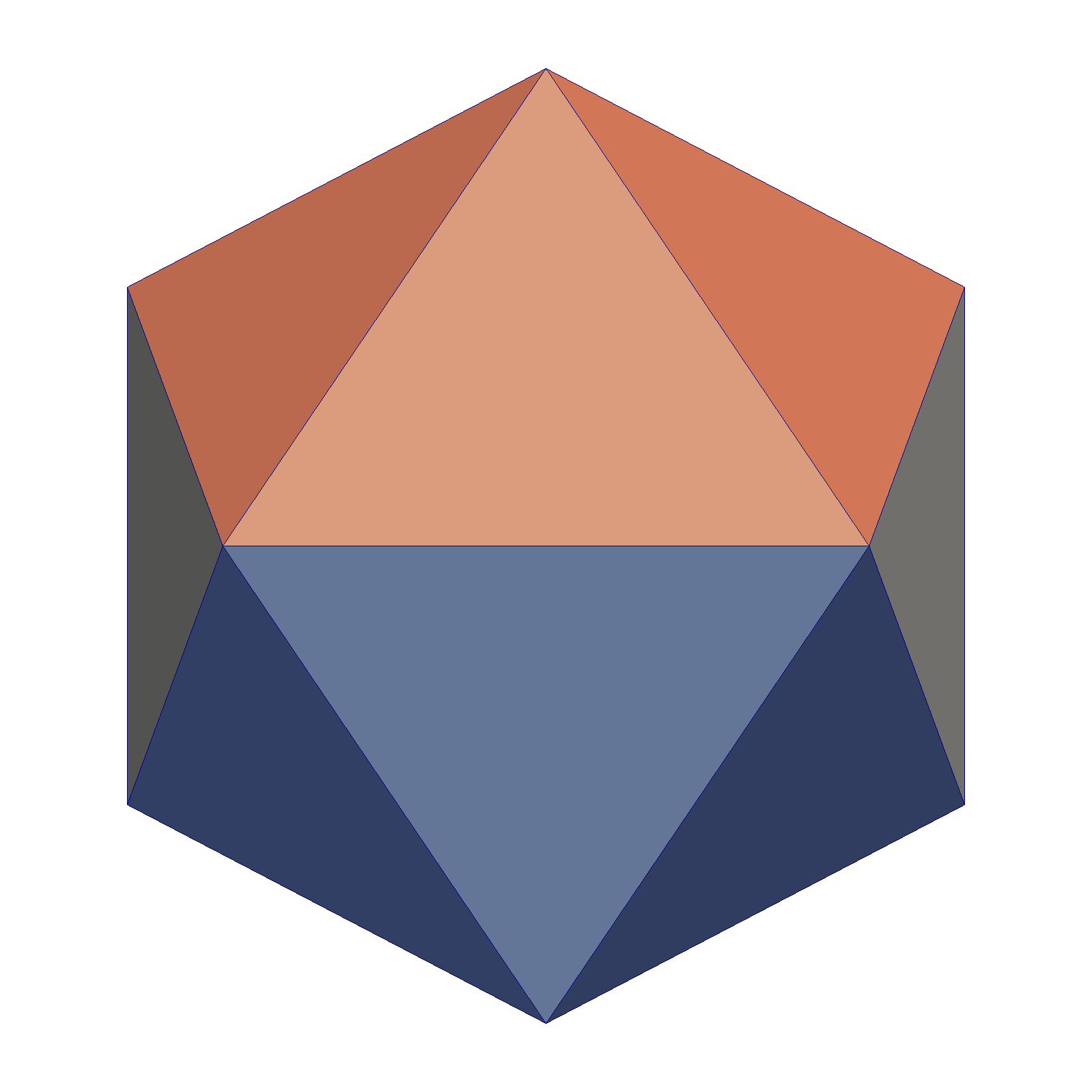} &
    \includegraphics[width=0.33\textwidth]{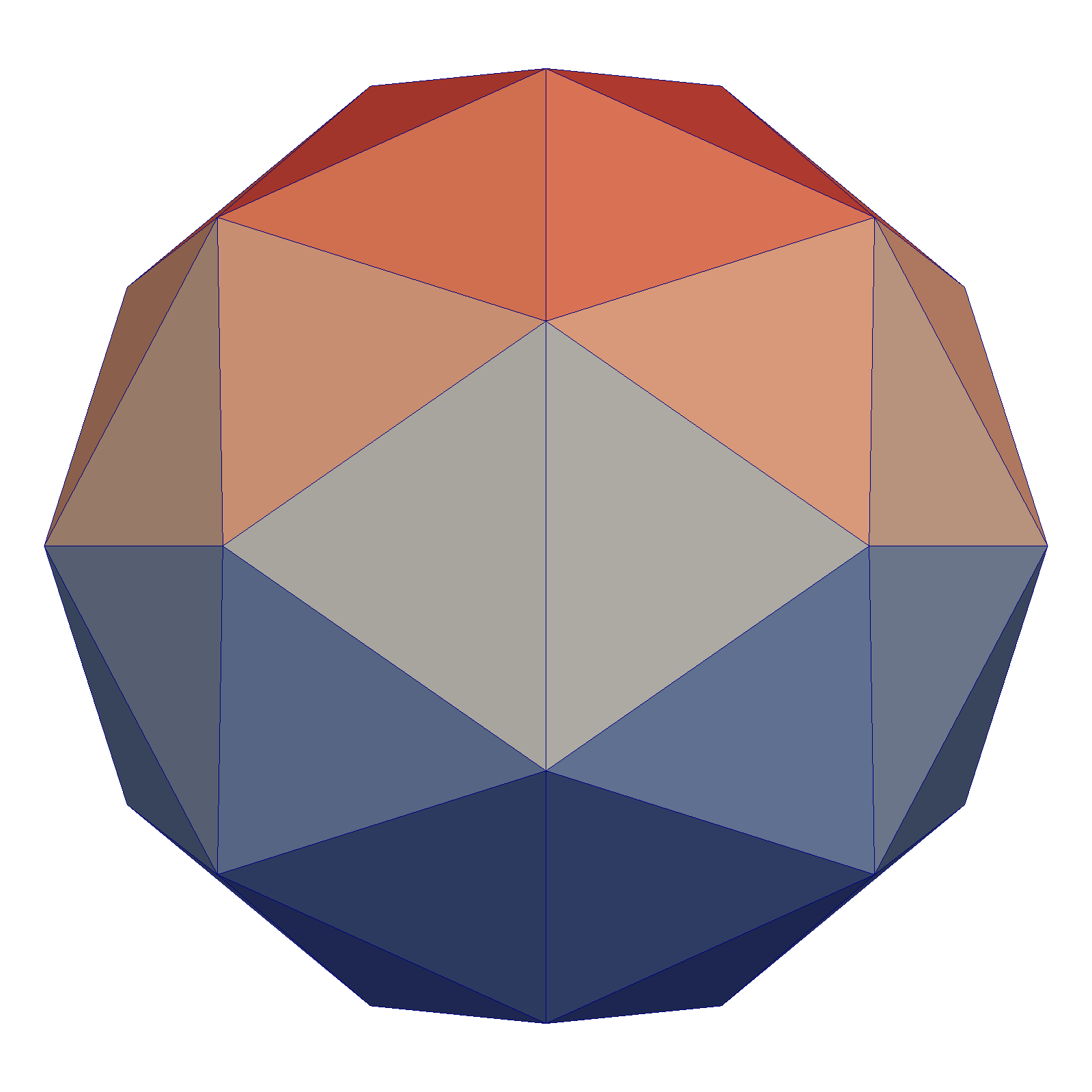} &
    \includegraphics[width=0.33\textwidth]{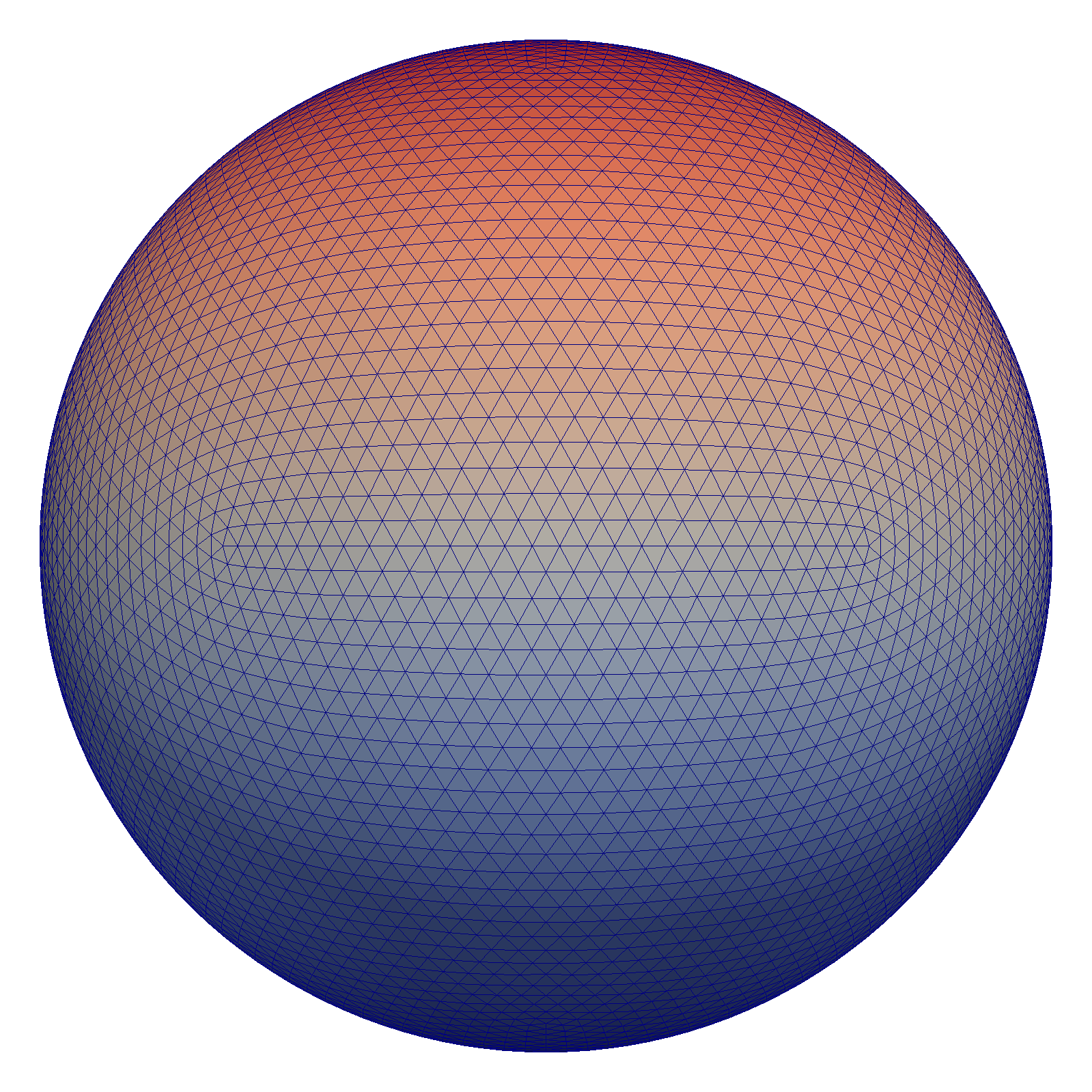} \\
    \begin{minipage}{0.33\textwidth}(a) Lowest resolution\end{minipage} &
    \begin{minipage}{0.33\textwidth}(b) One iteration of uniform refinement\end{minipage} &
    \begin{minipage}{0.33\textwidth}(c) Six iterations of uniform refinement\end{minipage}
  \end{tabular}
  \caption[Uniform refinement of a triangulation over the unit sphere]{
    Varying resolutions of a triangulation over the unit sphere used
    during initial data generation.
    The lowest resolution has $12$ vertices and $20$ triangles evenly
    distributed over the sphere.
    One iteration of uniform refinement leads to a triangulation with $32$
    vertices and $60$ triangles.
    Six iterations results in $7,292$ vertices.
  }
  \label{figUniformRefinement}
\end{figure*}

To establish an initially evenly distributed set of generators, we begin with
a regular icosahedron inscribing a unit sphere as seen in
the first panel of \cref{figUniformRefinement}.
This corresponds to our base resolution with $12$ vertices and $20$ triangles.
The triangular faces of the icosahedron form a triangulation over the sphere,
where each vertex corresponds to one generator of the
event horizon.
We will see later in \cref{secInitialData} exactly how we map from
this sphere to event horizon generators, but for now consider this to
closely represent the distribution of generators over an event horizon.

We can reach arbitrarily high resolutions by applying the following
triangle refinement procedure to each of the $20$ triangles on the surface:
\begin{enumerate}
  \item Choose a point at the median of the vertices of the triangle
    to be refined.
  \item Move the point radially outward to the surface of the unit sphere.
  \item Convert the original triangle to three smaller triangles
    by connecting the new point with the vertices of the original triangle.
  \label{itemAddPoint}
  \item Check the Delaunay condition, described below,
    along all exterior edges of the new
    triangles and perform an edge flip if necessary.
  \label{itemEdgeFlip}
\end{enumerate}
When we apply this procedure to all the triangles, we call it
\textit{uniform refinement}.

To understand the Delaunay condition and edge flips, consider four points
connected to form the quadrilateral $\square ABCD$.
There are two ways to form a set of two triangles from this quadrilateral,
either by connecting $\overline{AC}$ to form $\triangle ABC$ and
$\triangle ACD$, or connecting
$\overline{BD}$ to form $\triangle ABD$ and $\triangle BCD$.
The pair of triangles with the largest minimum angle among the six
interior angles satisfies the Delaunay condition.
An edge flip is the name for the process of converting a pair of triangles
with a shared edge that fails the Delaunay condition into one that satisfies
the condition.
For example, we could ``flip the edge'' $\overline{AC}$ by removing
$\overline{AC}$ and replacing it with $\overline{BD}$.

There are two choices for how to calculate the interior angles of these
triangles,
since the triangle vertices live on a sphere.
The code can handle
treating the triangles as either flat or curved along the surface of the sphere.
We default to treating the triangles as curved when calculating angles,
but this difference becomes less important as the triangles get
sufficiently small.

One round of uniform refinement adds a vertex to each triangle, going from
an icosahedron with
$12$ vertices to a Pentakis dodecahedron with $32$ vertices shown in
panel~(b) of \cref{figUniformRefinement}.
This procedure can be repeated indefinitely, but we typically uniformly
refine the full triangulation six times, resulting in
$7,292$ vertices evenly distributed over the surface as shown
in panel~(c) of \cref{figUniformRefinement}.
In general, the $n$\ts{th} iteration of uniform refinement has $20\times3^n$
triangles and $2+10 \times 3^n$ vertices\footnote{
  Every iteration of uniform refinement adds one vertex per triangle in the
  triangulation, so we have $12 + 20 \sum_{i=1}^n 3^{i-1}$ vertices at
  the $n$\ts{th} level of refinement.
}.

While there are faster ways to generate uniform distributions
of vertices over the sphere,
the refinement method we use is general and can be used to adaptively
refine arbitrary regions of the sphere by only refining a subset of
the triangles,
a procedure we call \textit{selective refinement}.
In practice,
we typically do a pilot event horizon run using a uniform distribution
of $7,292$ generators to determine the set of triangles we
are interested in refining.
Then we add generators to only those triangles in the region of interest
and perform a second event horizon run.
Selective refinement is crucial for studying small-scale features
of the event horizon, such as the short-lived hole in a toroidal EH surface
as discussed in the companion paper~\cite{BohnTorus2016}.

We have control over multiple parameters to tune the selective refinement:
\begin{itemize}
  \item The \textit{refinement depth} parameter roughly controls how many points
    are added to the selected triangles.
  \item The \textit{refinement width} parameter controls how wide a region
    we are refining.
  \item We can control how many event horizon iterations we perform.
\end{itemize}
The refinement depth and width provide complete control over the refinement
for the problems we are interested in,
so we usually set the number of EH runs to two, corresponding
to one round of refinement.

\begin{figure}
  \centering
  \begin{tabular}{cc}
    \includegraphics[width=0.4\columnwidth]{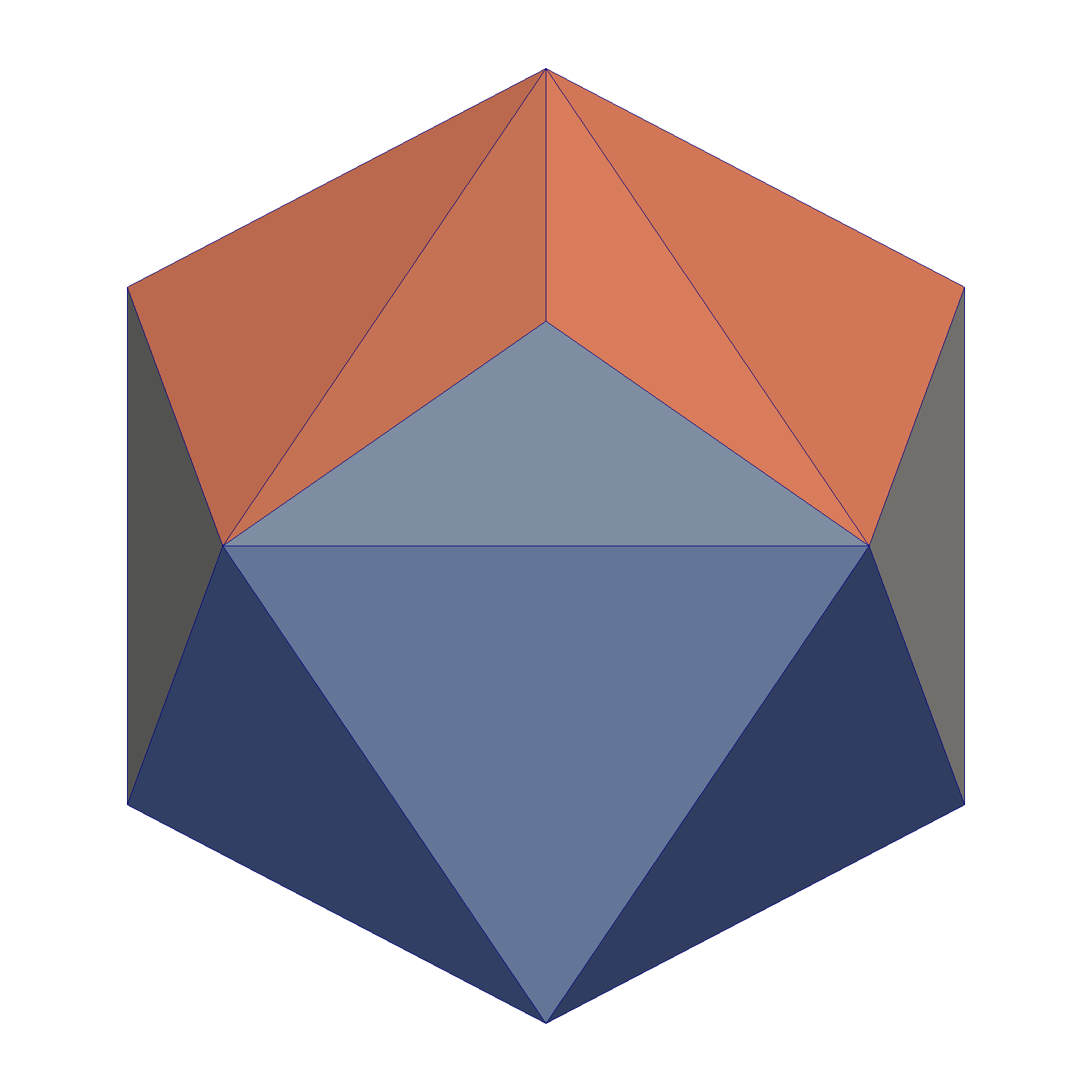} &
    \includegraphics[width=0.4\columnwidth]{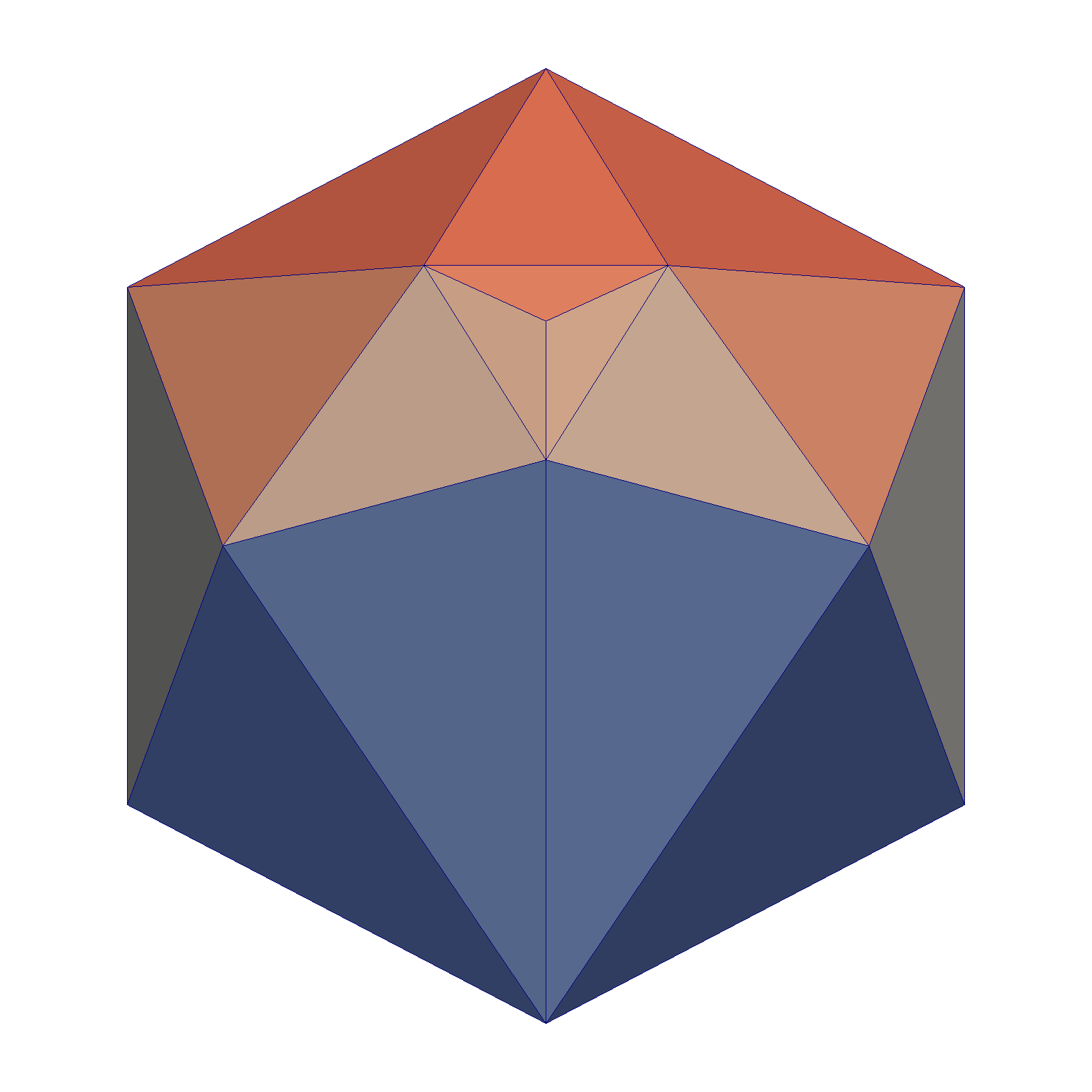} \\
    (a) One iteration &
    (b) Two iterations \\
    \includegraphics[width=0.4\columnwidth]{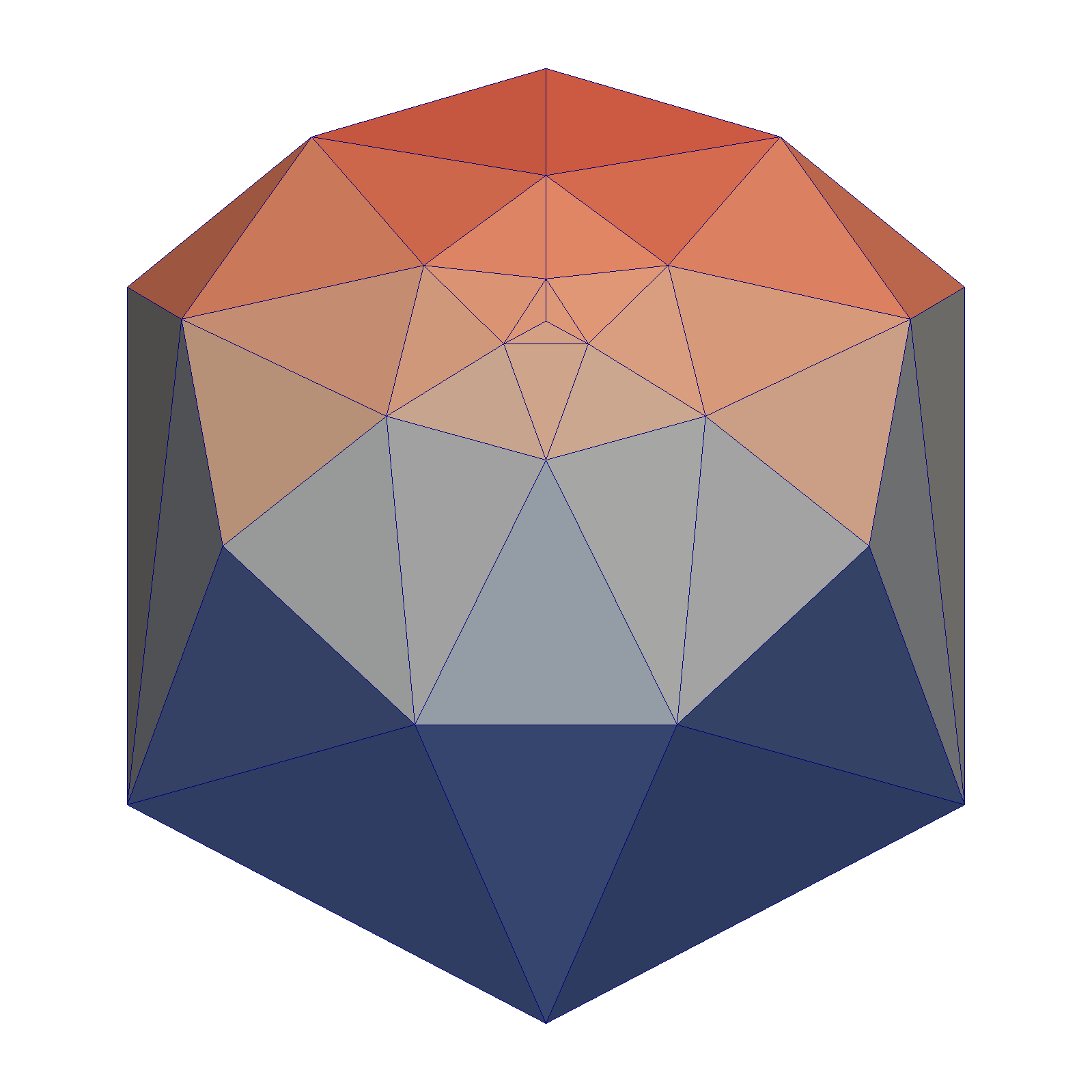} &
    \includegraphics[width=0.4\columnwidth]{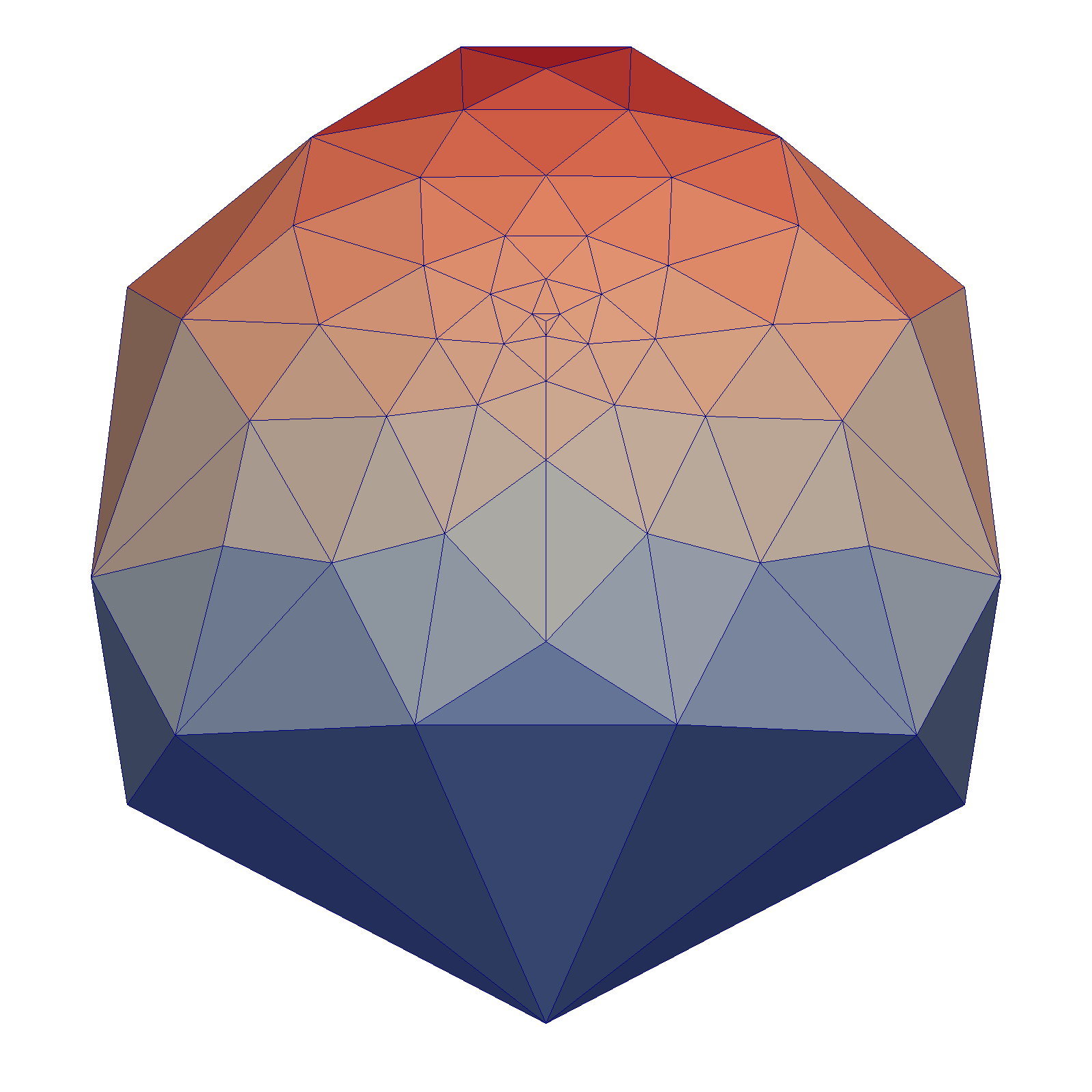} \\
    (c) Three iterations &
    (d) Four iterations
  \end{tabular}
  \caption[Selective refinement of one triangle of triangulation over a sphere]
  {Selective refinement of one triangle in the original $12$ vertex
    triangulation shown in \cref{figUniformRefinement}.
    Panels~(a)-(d) show
    one to four iterations of our
    refinement procedure applied to one triangle.
  }
  \label{figSelectiveRefinement}
\end{figure}

Before seeing examples of localized refinement,
we must introduce the concept of a triangle \textit{descendant}.
When refining one triangle, we add a vertex and convert the triangle
to three
new triangles that are all labeled descendants of the original triangle.
In addition, if we have to perform any edge flips, we convert two triangles
into two new triangles that
are both labeled descendants of the two previous triangles.
We maintain a full tree-like structure of triangles that is useful
for quickly locating triangles given a location on the sphere, but
more importantly the tree is useful when adding more than one point
to a triangle.

An example of selective refinement is shown in
\cref{figSelectiveRefinement}, where we explore
aggressive refinement of one triangle.
Panel~(a) shows one refinement iteration applied to
one triangle, where a
point
is added and connected to the vertices of the triangle.
The Delaunay condition is checked on all $3$ edges opposite the new vertex,
but in this instance, no edges needed to be flipped.
In panel~(b), to reach a second refinement iteration
we add a point to each of the three previously created
triangles, resulting in a total of $4$ new points.
In other words, we add a vertex to each descendant of the original triangle.
Again the Delaunay condition is checked on the edges opposing any of the
new vertices, which is $6$ edges in this case.
We can see that all $6$ edges are flipped here, giving an improved set of
triangles.
To perform a third refinement iteration, we must again add one vertex
to each of the $12$ descendant triangles of the original triangle
and check for edge flips.
The refinement depth is closely related to the number of refinement iterations.
Our highest resolution event horizon run to date refined from $7,292$
to $246,687$ generators with this procedure, and the algorithm handles
this with no problems.

Performing edge flips continually as we refine is important because
we add points to the median of each triangle.
If we want an even distribution of vertices, then we want each triangle to be
as close
to equilateral as possible, which amounts to maintaining a Delaunay
condition on the sphere.
These edge flips allow the density of vertices to change smoothly even though
there is a large range of vertex
densities over the sphere, as seen by comparing the density of vertices
in panel~(d) of \cref{figSelectiveRefinement} to the original vertices
in \cref{figUniformRefinement}.
In practice, the refinement does not stray far beyond the region
where we are interested in refining.

\begin{figure}
  \centering
  \includegraphics[width=0.8\columnwidth]{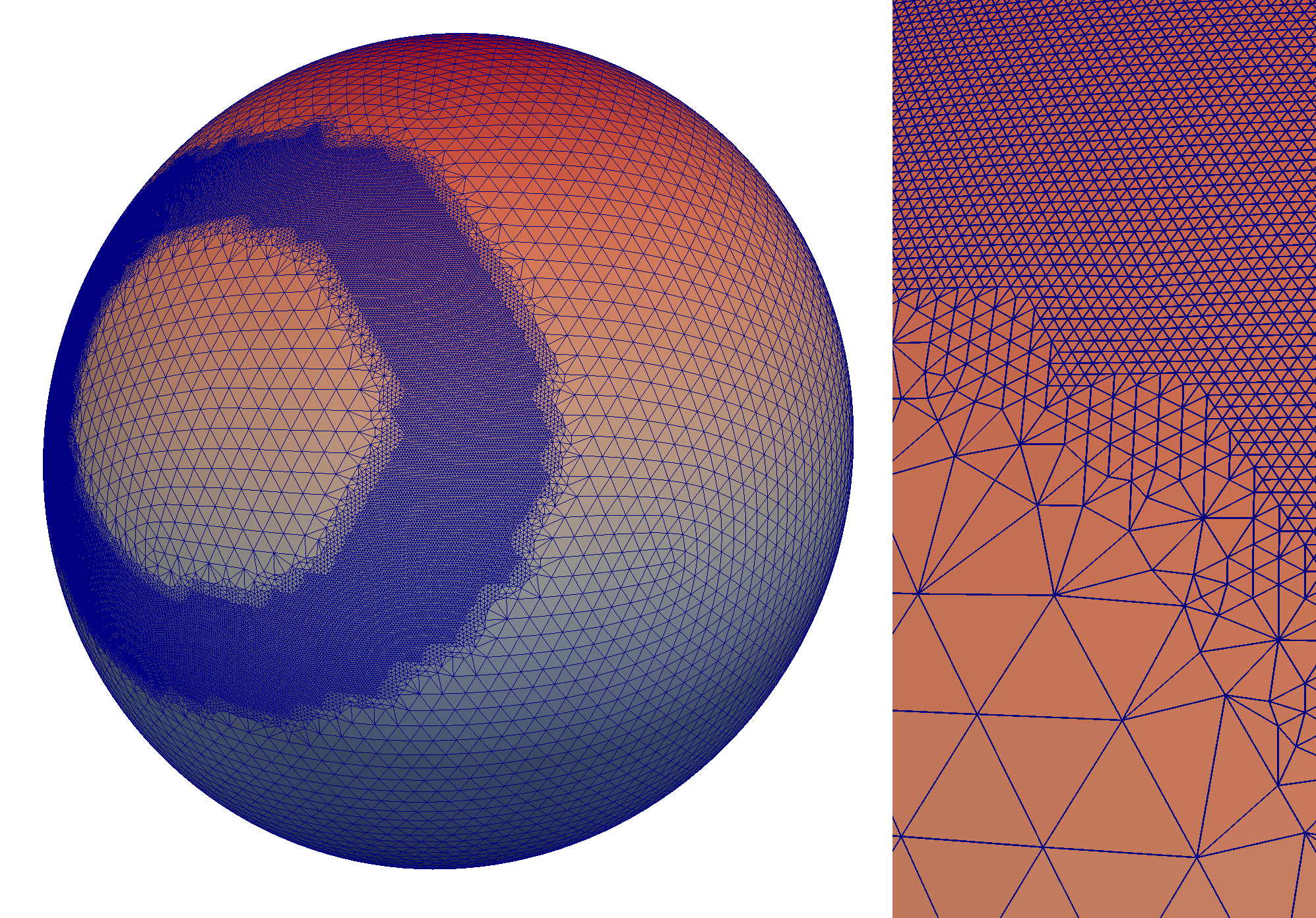}
  \caption[Event horizon generator initial data after selective refinement]{
    Selective refinement of event horizon generators for a BBH with mass ratio
    $6$,
    refining from $7,292$ generators to $49,350$ generators.
    The right section of the figure shows a zoomed-in region of the left
    section,
    highlighting the smooth transition of generator
    density over the initial data surface.
    The regions where refinement occurs are chosen to be around the
    generators associated with the neck of the event horizon during
    the BBH merger, as seen in
    \cref{figUltimateIDMerger}.
  }
  \label{figUltimateID}
\end{figure}

\begin{figure}
  \centering
  \includegraphics[width=0.8\columnwidth]{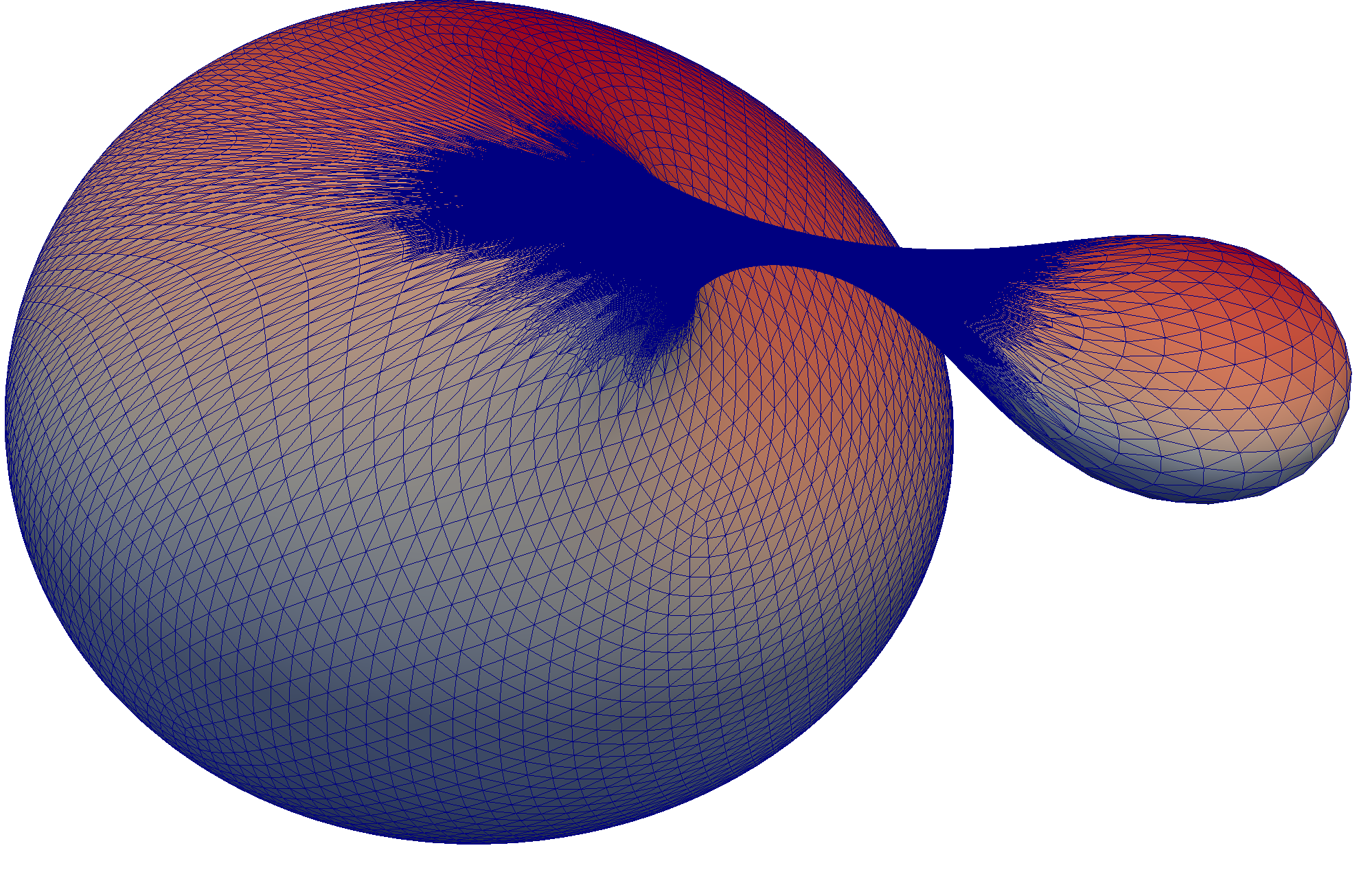}
  \caption[Refined generator locations just after a BBH merger]{
    Generator locations during the merger of a $6$ to $1$ mass ratio binary,
    for which initial data is shown in \cref{figUltimateID}.
    Refinement occurs in the neck of the event horizon, where future generators
    joined the event horizon surface earlier in the merger.
  }
  \label{figUltimateIDMerger}
\end{figure}

\Cref{figUltimateID} shows an example of selective refinement of
an event horizon surface for a binary.
The surface is the initial data surface for an event horizon simulation of
a $6$ to $1$ mass ratio binary, with dimensionless spin $\chi \approx 0.9$ on
the large BH and $\chi \approx 0.3$ on the small BH in arbitrary
directions~(\cite{SXSCatalog}~ID~SXS:BBH:0165).
To study the small scale features that arise where the event horizons
first touch,
we need to add generators to that portion of the surface.
The right side of this figure shows a zoomed-in region of the event
horizon surface to illustrate the transition between the low resolution
and high resolution regions.
\Cref{figUltimateIDMerger} shows the same simulation during the merger,
where we can see the high density of event horizon generators located
in the neck of the event horizon where the black holes met slightly
earlier.

The density of generators is smooth and continuous between the low
density and high density regions of generators.
This good behavior arises partially from continually checking the Delaunay
condition, as seen in \cref{figSelectiveRefinement}.
In addition, the code sets the number of refinement iterations to smoothly
transition between the low and high density regions automatically.
While our selective refinement algorithm refines triangles, we
determine which triangles to refine based on whether the generators
at the vertices of the triangle were future generators in the past.
If only one vertex of a triangle satisfies this property,
then we set the number of refinement iterations to
the specified refinement depth.
For each additional vertex of that triangle associated with the neck region,
we increment the number of refinement iterations by one.
For example, if the refinement depth parameter is set to $3$, as in
\cref{figUltimateID} and \cref{figUltimateIDMerger}, then we refine
triangles
along the border of the refinement region $3$ or $4$ iterations, and triangles
in the interior $5$ iterations.

The other tunable parameter is the refinement width, which controls
how wide our refinement region is.
Using \cref{figThorneSuggestion} as a reference, if we refined
triangles associated with the future generators in panel~(b) we would
obtain a fairly thin refinement region,
but if we refine based on the future generators in panel~(a) we would
widen the refinement region.
Therefore we control the refinement width by choosing how long before
the merger we identify triangles associated with future generators.

The parameters discussed so far refine the neck of the event horizon
satisfactorily,
but refine nowhere else.
For high mass ratio binaries, such as the one shown
in \cref{figUltimateIDMerger}, it may be worthwhile to consider refining
the surface based on the curvature as well.
In this case, the smaller black hole would have a large curvature compared
to the number of generators in the region,
and thus would have more generators added to that region.
One way to accomplish this refinement is to look at the angle between
the normal of a triangle and the normal of all its neighboring
triangles, and add generators if the angle is too large.
This type of refinement is not currently implemented since we are only
interested in the neck region for this paper.

After assembling a useful distribution of generators on the EH, how
do we calculate quantities over the surface?
Derivatives of scalars are calculated
using first-order finite differencing,
following~\cite{Thacker1977} adapted to a curved surface.
For some scalar $f$ defined at the vertices of the triangulation,
we can approximate the derivatives of the scalar inside the triangle
using the function values at the vertices
\begin{subequations}
\begin{align}
\label{eqTriDeriv}
  \partial_\theta f &\approx \left[ (\phi_2 - \phi_3)f_1 + \
    (\phi_3 - \phi_1) f_2 + \
    (\phi_1 - \phi_2) f_3 \right] / \Delta \\
  \partial_\phi f &\approx -\left[ (\theta_2 - \theta_3) f_1 + \
    (\theta_3 - \theta_1) f_2 + \
    (\theta_1 - \theta_2) f_3 \right] / \Delta \\
  \Delta &= (\phi_2 - \phi_3) \theta_1 + \
    (\phi_3 - \phi_1) \theta_2 + \
    (\phi_1 - \phi_2) \theta_3,
\end{align}
\end{subequations}
as in equation~$(1)$ of~\cite{Thacker1977}, where $f_i$ is the scalar value
at the vertex with coordinates $(\theta_i, \phi_i)$,
$\partial_{u} f$ is the partial
derivative of $f$ with respect to $u$, and $\Delta$ is twice the
coordinate area of the triangle.
To evaluate the derivative at a vertex, we perform a weighted average
of \cref{eqTriDeriv} over each triangle the vertex belongs to.
The derivative of the scalar at a vertex can thus be approximated as
\begin{subequations}
\begin{align}
\label{eqTriDerivVertex}
  \partial_\theta f &\approx \sum_{i=1}^{N} \
    (\phi_{i+1} - \phi_{i-1}) f_i / \Delta \\
  \partial_\phi f &\approx \frac{1}{\sin{\theta}} \sum_{i=1}^{N} \
    (\theta_{i+1} - \theta_{i-1}) f_i / \Delta \\
  \Delta &= \sum_{i=1}^{N} (\phi_{i+1} - \phi_{i-1}) \theta_i.
\end{align}
\end{subequations}
The derivatives in \cref{eqTriDerivVertex} are well-behaved far from the poles
of the $(\theta, \phi)$ coordinate system,
but would require care when crossing the poles.
Since our choice of vertices is evenly spread over the sphere, the vertices
do not avoid
the coordinate singularity at these poles.
To obtain well-behaved derivatives everywhere,
we set up three $(\theta, \phi)$ coordinate systems on the sphere with
the poles on the $x$, $y$, or $z$ axis, using a cyclic permutation of the
usual Cartesian to spherical coordinate transformation.
The derivative at some vertex uses all neighboring vertices, so the
lowest resolution triangulation our code supports
must be oriented such that
each vertex and its neighbors live entirely in one of the three coordinate
systems without crossing the poles.
This corresponds to vertices at all cyclic permutations of
$(\pm \phi, \pm 1, 0) / (\sqrt{1 + \phi^2})$, where $\phi$ is the golden ratio,
and we have normalized to $1$.
When computing quantities that do not depend explicitly on the choice
of coordinate system on the sphere, we simply choose the coordinate system
farthest
from the poles, i.e. $\theta$ closest to $\pi/2$.

\section{Generator evolution}
\label{secEvolution}

Our code can trace event horizon generators independently
through either numerical or analytic metric data,
which is useful for performing code tests.
It is common for numerical simulations
to use the $3+1$ decomposition~\cite{ADM},
so we
express the metric in the form
\begin{equation}
  \label{eqn:Decomposition}
  ds^2 = -\alpha^2 dt^2 + \gamma_{ij}(dx^i + \beta^i dt) (dx^j + \beta^j dt),
\end{equation}
where $\alpha$ is the lapse function, $\beta^i$ is the shift vector,
and $\gamma_{ij}$ is the spatial metric.\footnote{
  Our convention is that
  Greek indices, as in $x^\tau$,
  denote temporal or spatial components, while Latin
  indices, as in $x^i$, denote only spatial components.}
We obtain numerical data from simulations performed using the
Spectral Einstein Code
(\SpEC{})~\cite{SpECwebsite,
Szilagyi:2009qz, Hemberger:2012jz, SXSCatalog}.
The generators are traced by evolving a solution to
the geodesic equation
\begin{equation}
\label{eq:nullGeodesic}
\frac{d^2 x^\tau}{d\lambda^2} + \Gamma\indices{^\tau_\mu_\nu} \frac{dx^{\mu}}{d\lambda} \frac{dx^{\nu}}{d\lambda} = 0,
\end{equation}
where $x^\tau$
is the four-position of the geodesic, $\lambda$ is an affine
parameter, and $\Gamma\indices{^\tau_\mu_\nu}$ are the
Christoffel symbols describing the effective force
caused by spacetime curvature.

To facilitate the numerical geodesic evolution,
we split this second-order differential equation into two
first-order equations using an intermediate
momentum-like variable such as $p^\tau = dx^\tau / d\lambda$.
As we have some freedom in the definition of this momentum variable,
we look for one
that helps to
minimize computational time and numerical errors
when evolving through spacetimes with black holes.

We initially explored using the variable
$p_\tau = dx_\tau / d\lambda$ from Hughes~\textit{et al.}~\cite{Hughes1994},
along with converting
the evolution equations from affine parameter $\lambda$ to the coordinate time
$t$ of \SpEC{} evolutions through the use of $p^0 = dt/d\lambda$.
Although the resulting evolution equations are
concise and have no time
derivatives of metric variables,
the quantities $p^0$ and $p_i$ grow
exponentially near black hole horizons in
typical coordinate systems used by \SpEC{}
simulations.  This forces our time-stepper to take
prohibitively small steps in order to achieve the desired accuracy.

We therefore choose a momentum variable slightly different than $p_\tau$
to mitigate this time-stepping problem.
Null geodesics satisfy $\vec{p} \cdot \vec{p} = 0$,
which can be rewritten as
\hbox{$p^0=\alpha^{-1}(\gamma^{ij} p_i p_j)^{1/2}$} using the
metric in \cref{eqn:Decomposition}.
This expression shows that $p^0$ and $p_i$ scale similarly, so we can eliminate
the exponential behavior of these variables by evolving the ratio.
Our intermediate variable thus becomes
\begin{equation}
  \label{eqn:PiDefinition}
  \Pi_i \equiv \frac{p_i}{\alpha p^0} = \frac{p_i}{\sqrt{\gamma^{jk} p_j p_k}},
\end{equation}
where we also divide by $\alpha$ to reduce the number of terms
in the resulting evolution
equations.
This choice of intermediate variable is the same one that appears
in~\cite{Bohn2015}.
Using $\Pi_i$ and the $3+1$ decomposition of \cref{eqn:Decomposition},
we can express the geodesic equation in \cref{eq:nullGeodesic}
in the form
\ifdefined\conditional \else \begin{widetext} \fi
\begin{subequations}
\begin{align}
\frac{d \Pi_i}{dt} = {}& -\alpha_{,i}
    + (\alpha_{,j} \Pi^j - \alpha K_{jk} \Pi^j \Pi^k) \Pi_i
    + \beta\indices{^k_{,i}} \Pi_k
    - \frac{1}{2}\alpha \gamma\indices{^j^k_{,i}} \Pi_j \Pi_k \\
    \frac{ dx^i}{dt} = {}& \alpha \Pi^i - \beta^i,
\end{align}
\label{eqnEvEqns}
\end{subequations}
\ifdefined\conditional \else \end{widetext} \fi
where $K_{jk}$ is the extrinsic curvature (see, e.g.,~\cite{ADM}) and
$\Pi^i$ is defined via the inverse spatial metric as
$\Pi^i \equiv \gamma^{ij} \Pi_j$.
Note that the geodesic equation
consists of four second-order equations, yet we only
have three pairs of coupled first-order equations in \cref{eqnEvEqns}.
Because we are evolving a normalized momentum, \cref{eqn:PiDefinition},
we have lost information about $p^0$ during evolution.
Compared to the evolution equations in Hughes \textit{et al.}~\cite{Hughes1994},
we have introduced a time derivative of the
three-metric inside $K_{jk}$, but we
have significantly sped up the evolution near
black holes by removing the exponential growth of $p^0$ and $p_i$.

The equations in \cref{eqnEvEqns} are similar to those in~(28) of
Vincent \textit{et al.}~\cite{Vincent2012}.
In fact our intermediate evolution variable $\Pi_i$ is related to their
variable $V^i$ by the three-metric,
such that $\Pi^i = V^i$.
But our \cref{eqnEvEqns} has a
reduced number of both temporal and spatial derivatives of metric
quantities compared to Vincent's~(28).
More detailed information about splittings of the geodesic evolution equation
can be found in \cref{appendix:EvolutionEquations}.

\section{Handling metric data}
\label{secInterpolation}

Because we perform the generator evolution through
the \SpEC{} metric data backwards in time, we must complete
the binary black hole simulation beforehand while saving sufficient metric
data to disk.
We need all the metric components
specified in \cref{eqnEvEqns} at any given time and location in the
evolution domain, or we need to be able to compute them.
While we do not need all of the metric and its derivatives in our
evolution equations, it is
simpler to save $g_{\mu \nu}$ and all of the derivatives used during
the \SpEC{} BBH simulation and deal with slightly more disk space usage.

The metric and derivatives are stored on the BBH evolution grid points
at a deterministic set of times such that we can interpolate the metric
quantities to any spacetime point in the simulation domain.
The metric $g_{\mu \nu}$ has $10$ unique components when
accounting for symmetry, and the derivatives $\partial_\delta g_{\mu \nu}$
have $40$ components leading to a total of $50 N_{\rm{pts}} N_{t}$ numbers,
where $N_{\rm{pts}}$ is the average number of grid points and $N_{t}$
is the number
of time slices stored.
In addition, some extra information about where the points are located
and how they are distributed must also be stored.

For one fully generic BBH evolution of unequal mass black holes with
arbitrary spin directions and magnitudes,
the metric data can take many terrabytes of disk usage.
Since typical clusters have one or two gigabytes of memory per core, we
do not have nearly enough memory to read all the metric data at once.
To handle this situation, we utilize a shared memory paradigm by using OpenMP.
During generator evolution, we read sections of the metric data into
memory only as needed and at most once,
storing it in a shared thread-safe cache.
Other generators then simply access the cache to get the metric data instead of
reading it from disk for themselves.

We maintain a priority queue of generators ordered
by their current evolution time, such that
generators that are farthest behind are given highest priority.
After a pool of OpenMP threads is spawned, each thread will grab the
next highest priority generator in the queue, evolve for one timestep,
then insert the generator back into the priority queue.
A potential concern that the CPU cache was not being utilized
by taking only one timestep at a time turned out not to be valid.
With the priority queue, generators are kept as close in time as possible,
so that
metric data in the cache is kept for as little time as needed.
Since the domain structure in \SpEC{}
consists of many subdomains, only the
required subdomains are read into memory.
Periodically, we use the evolution time of the farthest-behind
generator to determine which metric data stored in the cache is safe
to be deleted\footnote{Given that the farthest-behind generator is at time $t$,
  determining which metric data times are safe to delete is more complicated
  than just comparing the stored times against $t$.
  This is because we need to perform time interpolation, so
  the interpolation stencil width is also a factor.}.

When a generator requests metric data at a particular location and time,
we must perform both a
spatial and a temporal interpolation in general.
Spatial interpolations are performed spectrally, taking advantage of
the pseudo-spectral grid used during \SpEC{} simulations.
We are left with the innocent looking tasks of temporal interpolation
and how to properly
combine temporal and spatial interpolations.
These tasks turn out to be quite complicated and are described in
\cref{secInterpolationDetails}.

\section{Initial data}
\label{secInitialData}

We evolve a set of event horizon generators backwards in time to
trace the event horizon surface, so we need to set an initial time,
location, and direction for each generator.
As hinted at by \cref{figThorneSuggestion}, the apparent horizon
and event horizon surfaces asymptotically approach each other after the merger.
If we set the initial time of the backwards evolution to be late enough,
the black hole will have settled to a nearly stationary solution
and the apparent horizon surface could be used as initial data for
the locations of the event horizon generators~\cite{Anninos1995}.
In \SpEC{}, the apparent horizon is represented with a spherical harmonic
decomposition, so we simply look for a time where the spherical harmonic
coefficients are sufficiently stationary to choose an initial time.

Next we need to determine the positions of the generators
using the triangulation over the unit sphere described in
\cref{secTriangulation}.
We first note that each vertex of the unit sphere triangulation defines
a ($\theta$, $\phi$) direction.
The position of the generator associated with that vertex is then set to
the intersection of the AH surface and the ray starting
at the center of the AH pointing in the direction
defined by the vertex.
We use spectral interpolation on the spherical harmonic basis used to
represent the AH to find the intersection.
Since stationary black hole AHs have a nearly spherical shape
when represented in typical coordinate systems used by \SpEC{},
mapping between the reference sphere and the AH surface
roughly maintains the carefully constructed distribution of vertices from
\cref{secTriangulation}.

Finally, we need to find the initial direction of each generator, used
to calculate our intermediate evolution variable $p_i / (\alpha p^0)$
from \cref{eqnEvEqns}.
Following~\cite{CohenPfeiffer2008}, the initial direction of a generator
should be the normal to the
surface at the location of the generator, where
the normal is calculated spectrally on the AH following
Baumgarte~\textit{et al.}~\cite{baumgarte_etal96}.
The normal direction is set to $p^i$, which is transformed into
$p_i / (\alpha p^0)$ using the lapse and
$p^0$ as calculated in \cref{secEvolution}.

It is important to note that refinement of the unit sphere in
\cref{secTriangulation} never destroys vertices, but only destroys
(and then creates) triangles.
Once we trace an EH generator trajectory, we can store and reuse the
trajectory after refinement without retracing the generator.
Therefore we only calculate initial data for newly created vertices
in the triangulation for which we need to find the trajectory.
Unfortunately, while the generator trajectories from the pilot run do not
need to be recalculated, determining when generators join the horizon
must be recalculated completely since the triangles have changed.

\section{Identifying future generators}
\label{secIdentifyingGenerators}

Although the event horizon surface is generated by null geodesics that never
leave the horizon, event horizon generators readily join onto the horizon
during the merger, as can be seen in \cref{figThorneSuggestion}.
In the backwards in time language, generators can leave the horizon
where they meet other generators through one of two types of points:
caustics, where neighboring generators converge to a point,
or crossover points, where non-neighboring generators on the horizon
meet.
We must therefore identify and distinguish these caustics and crossover points.

When we trace event horizon generators, we record their locations
at a predetermined set of times.
In order to properly resolve the short-duration features appearing during
the merger of the black holes, we need fine time resolution during
the merger.
However, the process of looking for caustics or crossover points scales
linearly with the number of times where we record generator locations.
We do not require such fine time resolution after merger where
the event horizon is slowly varying and no more generators are joining,
so we smoothly transition the separation
between recording times from the fine resolution merger to the coarse resolution
ringdown.
We use a piecewise function with a hyperbolic tangent transition function
to specify the spacing between recording times $\Delta t$,
\ifdefined\conditional
\begin{equation}
  \Delta t(t) =
\begin{cases}
  \hfill \Delta t_{\rm{coarse}} \hfill & \text{ $t_{\rm{coarse}} \le t$ } \\
  \hfill \displaystyle \!\begin{aligned}
    \Delta t_{\rm{coarse}} +& \left(\Delta t_{\rm{fine}} - \Delta t_{\rm{coarse}}\right) \times \\
  &\left[0.5 \left(1 + \tanh{\left\{\tan{\left(\pi \left(1.5 - \frac{t - t_{\rm{begin}}}{t_{\rm{fine}} - t_{\rm{coarse}}}\right)\right)}\right\}}\right)\right]\end{aligned}
    & \text{ $t_{\rm{fine}} \le t < t_{\rm{coarse}}$ } \\
    \hfill \Delta t_{\rm{fine}} \hfill   & \text{ $t < t_{\rm{fine}}$ },
\end{cases}
\label{eqnRecordingTimes}
\end{equation}
\else
\begin{widetext}
\begin{equation}
  \Delta t(t) =
\begin{cases}
  \hfill \Delta t_{\rm{coarse}} \hfill & \text{ $t_{\rm{coarse}} \le t$ } \\
  \hfill \displaystyle
    \Delta t_{\rm{coarse}} + \left(\Delta t_{\rm{fine}} - \Delta t_{\rm{coarse}}\right) \times
  \left[0.5 \left(1 + \tanh{\left\{\tan{\left(\pi \left(1.5 - \frac{t - t_{\rm{begin}}}{t_{\rm{fine}} - t_{\rm{coarse}}}\right)\right)}\right\}}\right)\right]
    & \text{ $t_{\rm{fine}} \le t < t_{\rm{coarse}}$ } \\
    \hfill \Delta t_{\rm{fine}} \hfill   & \text{ $t < t_{\rm{fine}}$ },
\end{cases}
\label{eqnRecordingTimes}
\end{equation}
\end{widetext}
\fi
where $\Delta t_{\rm{fine}}$ and $\Delta t_{\rm{coarse}}$ specify
the fine and coarse spacings, $t_{\rm{fine}}$ and $t_{\rm{coarse}}$
specify the boundaries for the fine and coarse spacing regions,
and the transition function in square brackets varies between $0$ and $1$.
The time range between $t_{\rm{fine}}$ and $t_{\rm{coarse}}$ is used to smoothly
transition between the different spacings, and
any smooth monotonic transition function would be sufficient.

After performing the tracing, we must determine if and when generators leave
the horizon backwards in time using the recorded generator locations.
We search for caustics, where neighboring generators meet, by looking
for generators with negative expansion parameter,
similar to Cohen~\textit{et al.}~\cite{CohenPfeiffer2008}.
The expansion of a generator is proportional to the fractional change of
the area element around the generator,
\begin{equation}
  \label{eqnExpansion}
  \theta \propto \frac{1}{\sqrt{h}} \frac{\partial \sqrt{h}}{\partial t},
\end{equation}
where $h$ is the determinant of the induced metric on the horizon
at the location of the generator.
Unlike Cohen~\textit{et al.}~\cite{CohenPfeiffer2008},
where the induced metric is found
using second-order finite difference stencils,
it is not trivial to go beyond first-order finite differencing using our
irregularly structured grid.
Nevertheless, we
see no evidence that the first-order derivatives are not accurate enough,
since the
adaptive refinement scheme discussed in \cref{secTriangulation}
drastically decreases the distance between generators.

The induced metric on the event horizon is given by
\begin{equation}
  h_{ab} = \gamma_{ij} \frac{\partial q^i}{\partial y^a} \frac{\partial q^j}{\partial y^b},
\end{equation}
where $\gamma_{ij}$ is the spatial metric,
$q^i$ are the coordinates on the $3$-dimensional spacetime slice, and
$y^a$ are the $(\theta, \phi)$ coordinates on the horizon surface.
The derivatives are calculated using \cref{eqTriDerivVertex}.
Since we are only interested in the fractional change in $\sqrt{\det{h_{ab}}}$
in \cref{eqnExpansion},
we are free to perform a useful rescaling of the induced metric such that
\begin{equation}
  \tilde{h} = \det{\tilde{h}_{ab}} = \frac{1}{\sin^2\theta} \det{h_{ab}}.
\end{equation}
For a spherically symmetric space, $\tilde{h}$ is a constant
over the sphere, which
provides a useful correctness check and removes the
coordinate dependence on $\theta$.

When computing derivatives on the event horizon surface, to avoid
coordinate issues around the poles of the coordinate system, we can
align the poles with the $x$, $y$, or $z$ axes by choosing
the corresponding coordinate system defined in \cref{secTriangulation}.
We are free to change coordinate systems when calculating the expansion
for different generators since we are not comparing neighboring generators,
but only checking the sign of the expansion parameter.

To find the specific time $t_{\rm{join}}$ that a generator joins on the
horizon, we
first compute $\sqrt{\tilde{h}}$ for each generator at each stored time.
Then we take the partial derivative with respect to time along each generator
with a third order Lagrange interpolating polynomial
and calculate
the fractional change of $\sqrt{\tilde{h}}$ with respect to time,
which is proportional to the expansion parameter.
If this fractional change with respect to time changes sign between
two recording times, we know the join time is between these times.
We identify $t_{\rm{join}}$ by simply linearly interpolating
the fractional change between the recording times where it changes sign
to find when the expansion parameter passes through zero.

This algorithm to compute the expansion is parallelized using
a set of MPI processes and a pool of OpenMP threads on each process.
The set of generators on the event horizon surface is distributed
evenly across the OpenMP threads and MPI processes to calculate the quantity
$\sqrt{\tilde{h}}$.
The next step is to take the time derivative, which is a relatively
inexpensive operation, so it is currently only parallelized
over the MPI processes and not over OpenMP threads.

The other way generators can join the surface is through crossover points,
where non-neighboring generators meet.
Since we are evolving a finite number of generators to approximate the surface,
in general the generators we evolve will not cross each other.
We therefore look for crossover points by checking for surface
self-intersections by using a collision detection algorithm
as described
in~\cite{Cohen2012}, where every vertex is compared against
every triangle to see if the generator at that vertex passed through the
triangle between neighboring recording times.
Our situation is simplified compared to Cohen~\textit{et al.}~\cite{Cohen2012},
however, because
we explicitly start with an unchanging set of triangles
as opposed to needing to define and construct a set of triangles from a
$(\theta, \phi)$ grid.
Because the event horizon surface is approximated by connecting generators to
form
triangles, the collision of a triangle and any other generator
between times
$t_0$ and $t_1$ indicates that the generator joined the event horizon
through a crossover at a time $t_{\rm{join}}$ satisfying
$t_0 \le t_{\rm{join}} < t_1$.

\begin{figure}
  \centering
  \includegraphics[width=0.5\textwidth]{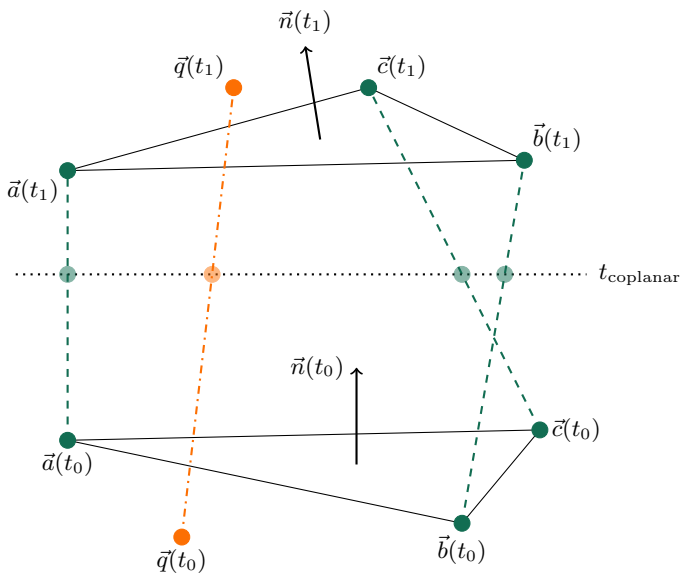}
  \caption[Schematic for collision detection between a moving triangle and
    line segment]
  {Checking for the collision of a moving triangle $\triangle abc$
    and some other generator $\vec{q}(t)$ between times $t_0$ and $t_1$.
    The triangle is constructed by connecting three neighboring vertices
    $\vec{a}(t)$, $\vec{b}(t)$, and $\vec{c}{t}$.
    At some time, the four generators may become coplanar, as illustrated
    in the figure.
  }
  \label{figCollisionDetection}
\end{figure}

The collision detection algorithm assumes each generator moves linearly
through time between two neighboring times, as shown in
\cref{figCollisionDetection}.
The location of a generator $\vec{q}(t)$ is therefore a linear function between
$t_0$ and $t_1$,
\begin{equation}
  \forall t \in [t_0, t_1), \hspace{10pt}
    \vec{q}(t) = \vec{q}(t_0)
    + \frac{t - t_0}{t_1 - t_0}\left(\vec{q}(t_1)-\vec{q}(t_0)\right),
\end{equation}
and similarly for the vertices of some triangle,
$\vec{a}(t)$, $\vec{b}(t)$, and $\vec{c}(t)$.
These generators and their trajectories are shown in
\cref{figCollisionDetection}.
The normal vector to the triangle $\triangle abc$ is then a quadratic
function in time
\begin{equation}
  \vec{n}(t) = \left(\vec{b}(t)-\vec{a}(t)\right) \times \left(\vec{c}(t) - \vec{a}(t)\right).
\end{equation}
We first solve for all times when the generator $\vec{q}(t)$ and the
triangle are coplanar by finding the roots of
\begin{equation}
  \vec{n}(t) \cdot (\vec{q}(t) - \vec{a}(t)) = 0,
\end{equation}
which is a cubic polynomial.
An example of a coplanar time is shown in \cref{figCollisionDetection}
as the dotted horizontal line.
We immediately disregard any roots of the cubic that lie outside the range
$t_0 \leq t_{\rm{coplanar}} < t_1$, and disregard complex roots.
For every root $t_{\rm{coplanar}}$ remaining, we check whether
$\vec{q}(t_{\rm{coplanar}})$ lies inside triangle $\triangle abc$.
If so, we mark the time at which the generator joins the horizon
as this time, $t_{\rm{join}} = t_{\rm{coplanar}}$.
It is possible that multiple roots of the cubic lie both in the desired
time range and inside the triangle, but generators physically
cannot cross after they join the horizon~\cite{MTW}, so we choose
the latest of the $t_{\rm{coplanar}}$ roots to be the join time.

We apply the collision detection algorithm for
every pair of neighboring times where we have recorded generator location data,
comparing each triangle to every other vertex.
Since all generators of the event horizon are on the event horizon surface
at late times,
we start with the latest pair of neighboring times and work backwards.
Since we are only interested in self-intersections of the actual event
horizon surface, we must remove vertices and triangles from the algorithm after
they leave the EH backwards in time.
Once we find a join time $t_{\rm{join}}$ for a generator, either corresponding
to joining as a caustic or a crossover, we do not
need to check for collisions with that generator and other triangles
as we move to earlier pairs of times where
$t_0 < t_1 < t_{\rm{join}}$.
Choosing when to remove a triangle from the algorithm is more subtle,
since triangles are formed from three EH generators.
We only remove a triangle from the algorithm once all three
generators forming the triangle have left the EH surface, a choice
that is described in \cref{appendixCollisionDetection}.

This includes both when the generator would be used to form a triangle as well
as when the generator would be the single vertex.
If both the caustic and crossover point algorithms
determine that $t_a \le t_{\rm{join}} \le t_b$ for some pair
of neighboring times $t_a$ and $t_b$, then the later time must be the
true join time, to satisfy the property that generators do not meet
after they join the event horizon.

The search for crossovers is the most costly part of the event horizon
simulation, since it is the only part of the simulation that scales
quadratically with the number of generators.
We have optimized the cost of each individual check for a collision between a
vertex and a triangle to $\mathcal{O}(2 \unit{\mu s})$.
In addition, for each pair of neighboring times, we use an OpenMP thread pool to
parallelize over the triangles, and we parallelize all the remaining
vertices over the MPI processes.

\section{Conclusions}
\label{secConclusions}

In this paper, we have presented a new event horizon finding code,
with adaptive localized refinement, based on a Delaunay triangulation on
a surface with the topology of a sphere.
We now have the ability to refine arbitrary portions of
the event horizon surface to discover and study small-scale features such as the
hole in a toroidal event horizon, as discussed in our companion
paper~\cite{BohnTorus2016}.
The triangulation is covered by three overlapping coordinate systems
to avoid issues with coordinate singularities at the poles of the standard
polar coordinate system.
Using
the backwards geodesic event horizon finding algorithm,
we specify how to calculate initial data for event horizon generators and
how to use the triangulation when searching for future
generators of the event horizon.

There are several ways this event horizon finding code can be further
improved.
The refinement algorithm currently creates an even distribution of
event horizon generators at late times in the BBH simulation, where the
horizon looks like Kerr.
Unfortunately, when traced backwards in time, the event horizon
surface becomes significantly stretched and distorted, leading
the triangles and the distribution of generators to be similarly stretched.
Since we are interested in studying the event horizon at the time of merger,
we would like the generators to be evenly spaced at the time of merger.
An improvement to the refinement algorithm would be to first perform an
event horizon run using an even distribution of generators to determine
how the triangles are stretched near merger,
then use the stretch information to add new
generators to the initial data surface so that the triangles are initially
stretched in the orthogonal direction, but become unstretched near the merger
into almost equilateral triangles.
It is not obvious to us how to generate such a distribution.
We note that it is difficult to re-triangulate the event horizon surface
at every time step, because the re-triangulation procedure would need
to understand that the surface is stretched, or else it would ``cut corners''
off the strongly distorted EH shape.

Furthermore, the collision detection algorithm,
the slowest step in the EH locating process,
is naively $\mathcal{O}(N^2)$ in the number of EH generators.
One could improve the coefficient of this algorithm by dividing
the space into spatial bins, with a quadtree for example,
and ignoring collisions of a triangle and generator
in entirely distinct spatial bins.
This was not implemented because of the complexity of determining a good
splitting of the surface and the problem of handling triangles or vertices that
move between different regions.

\appendix
\section{Null geodesic evolution equations in the \texorpdfstring{$3+1$}{3+1}
  decomposition}
\label{appendix:EvolutionEquations}

It is common for numerical simulations to use the $3+1$
decomposition~\cite{ADM}, so we express the metric \cref{eqn:Decomposition}
in the form
\begin{equation}
  g_{\mu \nu} =
  \left[
    \begin{array}{ccc}
      -\alpha^2 + \beta^i \beta_i & \gamma_{ij}\beta^i \\
      \gamma_{ij} \beta^j & \gamma_{ij}
    \end{array}
  \right].
\end{equation}

The inverse metric is
\begin{equation}
  g^{\mu \nu} =
  \left[
    \begin{array}{ccc}
      {\displaystyle -\frac{1}{\alpha^2}} & \
      {\displaystyle \frac{\beta^j}{\alpha^2}} \\ [1em]
      {\displaystyle \frac{\beta^i}{\alpha^2}} & \
      {\displaystyle \gamma^{ij} - \frac{\beta^i \beta^j}{\alpha^2}}
    \end{array}
  \right].
\end{equation}

The associated connection coefficients
for this representation of the metric are
\begin{equation}
\label{eqnAffineParameters}
\begin{split}
  \Gamma\indices{^0_0_0} =& \frac{1}{\alpha} \left(\alpha_{,t}
    + \beta^k \alpha_{,k} - K_{ij} \beta^i \beta^j\right) \\
  \Gamma\indices{^k_0_0} =& \gamma^{kj}\left(\beta_{j, t}
    + \alpha \alpha_{, j}
    - \frac{1}{2} \left(\gamma_{mn} \beta^m \beta^n\right)_{,j}\right)
    - \beta^k \Gamma\indices{^0_0_0} \\
  \Gamma\indices{^0_i_0} =&
    \frac{1}{\alpha}\left(\alpha_{,i} - K_{ij} \beta^j\right) \\
  \Gamma\indices{^k_i_0} =& -\alpha K\indices{_i^k}
    + \tensor[^{(3)}]{\nabla}{_i} \beta^k  - \Gamma\indices{^0_i_0}  \beta^k \\
  \Gamma\indices{^0_i_j} =&  -\frac{1}{\alpha} K_{ij} \\
  \Gamma\indices{^k_i_j} =& \tensor[^{(3)}]{\Gamma}{^k_i_j}
    + \frac{K_{ij}}{\alpha} \beta^k = \tensor[^{(3)}]{\Gamma}{^k_i_j}
    - \Gamma\indices{^0_i_j} \beta^k,
\end{split}
\end{equation}
where $\tensor[^{(3)}]{\nabla}{_i}$ and $\tensor[^{(3)}]{\Gamma}{^k_i_j}$ are
the covariant derivative and connection coefficients associated with the
spatial metric $\gamma_{ij}$, and we have used the extrinsic curvature
\begin{equation}
  K_{ij} = \frac{1}{2\alpha} \left(-\gamma_{ij,t}
  + 2\gamma_{ik} \beta\indices{^k_{,j}} + \gamma_{ij,m} \beta^m\right).
\end{equation}

To numerically integrate the geodesic equation
\begin{equation}
  \frac{d^2 x^\tau}{d\lambda^2} + \Gamma\indices{^\tau_\mu_\nu}
    \frac{dx^{\mu}}{d\lambda} \frac{dx^{\nu}}{d\lambda} = 0,
\end{equation}
we seek an efficient splitting into two first-order differential equations.
A natural splitting arises through the use of the photon momentum
\begin{equation}
  p^\mu = \frac{dx^\mu}{d\lambda}.
\end{equation}
With this momentum variable, we have the evolution equations
\begin{subequations}
\begin{align}
  \frac{dp^\tau}{d\lambda} &= -\Gamma\indices{^\tau_\mu_\nu} p^\mu p^\nu \\
  \frac{dx^\tau}{d\lambda} &= p^\tau.
\end{align}
\end{subequations}
These can be converted to equations with respect to a coordinate time $t$ by
diving through by $p^0 = dt/d\lambda$.

Cohen~\textit{et al.}~\cite{CohenPfeiffer2008} use a similar form by evolving
the quantity $p^i / p^0$
as an intermediate variable, although they define
the variable $p^i$ to be what is called $p^i / p^0$ here.
This intermediate variable gives the evolution equations
\begin{subequations}
\begin{align}
  \frac{d}{dt}\left(\frac{p^i}{p^0}\right) &=
    \left(\Gamma\indices{^0_\mu_\nu} \frac{p^i}{p^0}
      - \Gamma\indices{^i_\mu_\nu} \right) \frac{p^\mu}{p^0} \frac{p^\nu}{p^0}\\
  \frac{dx^i}{dt} &= \frac{p^i}{p^0},
\end{align}
\label{eqnCohenEquations}
\end{subequations}
which is a convenient intermediate variable choice as we will see shortly, but
is problematic
because it involves all of the connection coefficients during
evolution.
Additionally, the use of $\Gamma\indices{^\mu_0_0}$ involves time derivatives
of the lapse and shift (\cref{eqnAffineParameters}).

Performing the sum over all the connection coefficients is inefficient
because of the number of terms being summed as well as inaccurate if the
metric terms come from a numerical source versus an analytic source.
There are many cancellations in the geodesic equation that can be taken
advantage of with the appropriate choice of intermediate variable.
Hughes~et~al.~\cite{Hughes1994} explored using
\begin{equation}
  p_\mu = g_{\mu \nu} p^\nu,
\end{equation}
obtaining the evolution equations
\begin{subequations}
\begin{align}
  \frac{dp_i}{d\lambda} &= - \alpha \alpha_{,i} \left(p^0\right)^2
    + \beta\indices{^k_{,i}} p_k p^0
    - \frac{1}{2} \gamma\indices{^j^k_{,i}} p_j p_k \\
  \frac{dx^i}{d\lambda} &=  \gamma^{ij} p_j - \beta^i p^0.
\end{align}
\end{subequations}
Converting to an evolution with respect to coordinate time $t$ gives
\begin{subequations}
\begin{align}
  \frac{dp_i}{dt} &= -\alpha \alpha_{,i} p^0 + \beta\indices{^k_{,i}} p_k
    - \frac{1}{2} \gamma\indices{^{jk}_{,i}} \frac{p_j p_k}{p^0} \\
  \frac{dx^i}{dt} &= \gamma^{ij} \frac{p_j}{p^0} - \beta^i.
\end{align}
\label{eqnHughesEquations}
\end{subequations}
These equations have considerably fewer terms than those in
\cref{eqnCohenEquations} and also no time derivatives of metric functions.
We note that although the variable $p^0$ is not evolved, it can be calculated
by enforcing $\vec{p} \cdot \vec{p} = 0$, giving
$p^0 = \sqrt{\gamma^{ij} p_i p_j} / \alpha$.

Unfortunately, these equations are poorly suited for evolving outgoing null
geodesics
near black hole horizons in the coordinate systems we are interested in,
as $p^0~\sim~e^t$ for an event horizon generator of a Schwarzschild spacetime
expressed in Kerr-Schild coordinates for example.
Other components of the $4$-momentum have similar exponential dependence,
leading
to increasingly small timesteps.
The evolution equations in \cref{eqnCohenEquations}
conveniently cancel the exponential behavior by evolving the ratio
$p^i~/~p^0$.
Can we get the best of both worlds, avoiding the exponential behavior
of \cref{eqnHughesEquations} and avoiding the large number of terms
in \cref{eqnCohenEquations}?

One attempt is to evolve the lower momentum normalized by $p^0$ as in
\begin{equation}
  P_i \equiv \frac{p_i}{p^0}.
\end{equation}
With the definition $P^i = \gamma^{ij} P_j$, this yields
the evolution equations
\ifdefined\conditional \else \begin{widetext} \fi
\begin{subequations}
\begin{align}
  \frac{dP_i}{dt} =& -\alpha \alpha_{,i} + \beta\indices{^k_{,i}} P_k
    - \frac{1}{2}\gamma\indices{^j^k_{,i}} P_j P_k
   + \frac{P_i}{\alpha}\left(-\alpha_{,j} \beta^j
   + 2\alpha_{,j}P^j
    + \dot{\alpha} - K_{jk} P^j P^k\right) \\
  \frac{dx^i}{dt} =& P^i - \beta^i
\end{align}
\end{subequations}
\ifdefined\conditional \else \end{widetext} \fi
These equations certainly have more terms than \cref{eqnHughesEquations},
but do not suffer from the issue of small timesteps.

We can reduce the number of terms involved in the equations further
by including an extra factor of the lapse, such that
\begin{equation}
  \Pi_i \equiv \frac{p_i}{\alpha p^0} = \frac{P_i}{\alpha}
    = \frac{p_i}{\sqrt{\gamma^{jk} p_j p_k}}.
\end{equation}
Similarly, we define $\Pi^i = \gamma^{ij} \Pi_j$.
The resulting evolution equations are those mentioned in the main text, which
we repeat here for completeness
\ifdefined\conditional \else \begin{widetext} \fi
\begin{subequations}
\begin{align}
\frac{d \Pi_i}{dt} = {}& -\alpha_{,i}
    + \left(\alpha_{,j} \Pi^j - \alpha K_{jk} \Pi^j \Pi^k\right) \Pi_i
    + \beta\indices{^k_{,i}} \Pi_k
    - \frac{1}{2}\alpha \gamma\indices{^j^k_{,i}} \Pi_j \Pi_k \\
    \frac{dx^i}{dt} = {}& \alpha \Pi^i - \beta^i.
\end{align}
\end{subequations}
\ifdefined\conditional \else \end{widetext} \fi
By using the variable $\Pi_i$, we have reduced further the number of terms
involved, eliminated time derivatives of the metric,
as well as removed the small timestep behavior.
Since we are evolving a normalized momentum, we have lost the ability
to calculate $p^0$.
If $p^0$ is necessary it can be evolved separately, but for outgoing geodesics
near black hole horizons $p^0~\sim~e^t$.
For such geodesics, we recommend evolving the quantity
$\ln\left(\alpha p^0\right)$, giving
\begin{equation}
  \frac{d \ln(\alpha p^0)}{dt} = -\alpha_{,i} \Pi^i + \alpha K_{ij} \Pi^i \Pi^j.
\end{equation}
An alternative is to simply evolve $\ln\left(p^0\right)$,
which has more terms.
For geodesics evolved far from black hole horizons, $p^0$ can be
evolved directly by noting
$d\ln\left(p^0\right)/dt~=~\left(1/p^0\right)~\left(dp^0/dt\right)$.

The evolution equations using $\Pi_i$ are similar to those in Equation~(28) of
Vincent \textit{et al.}~\cite{Vincent2012}.
In fact, the intermediate evolution variable $\Pi_i$ is related to
their variable $V^i$ by the three-metric, such that $\Pi^i = V^i$.
But our \cref{eqnEvEqns} has a reduced number of both
temporal and spatial derivatives of metric
quantities compared to Vincent's Eq.~(28).

\section{Spacetime interpolations}
\label{secInterpolationDetails}

Each component of the metric is handled independently, so it is sufficient to
consider the interpolation of a scalar $A$ defined on a set of points split
into separate subdomains and on a set of time slices.
This is complicated by the fact that \SpEC{} utilizes a dual-frame
system~\cite{Scheel2006, Hemberger:2012jz},
where computations are performed in a reference frame
called the grid frame.
In the grid frame, the black holes are stationary with respect to the
collocation points of the evolution, and a time-dependent mapping is maintained
between this frame and the asymptotically inertial frame, which we call
the inertial frame.

\definecolor{myteal}{RGB}{27, 158, 119}
\definecolor{myorange}{RGB}{217, 95, 02}
\definecolor{mypurple}{RGB}{117, 112, 179}
\definecolor{mypink}{RGB}{231, 41, 138}
\tikzset{
  griddot/.style={
    draw,
    circle,
    minimum size=12pt,
    color=myteal,
  },
  interpdot/.style={
    draw,
    circle,
    minimum size=7pt,
    color=myorange,
    fill=myorange,
  },
  purpdot/.style={
    draw,
    circle,
    minimum size=7pt,
    color=mypurple,
    fill=mypurple,
  },
  gridsep/.style={
    draw,
    dashed,
    very thick,
    mypurple
  },
  gridchangedot/.style={
    draw,
    circle,
    fill=black,
    minimum size=5pt,
  },
  plus/.style={cross out, draw,
    minimum size=2*(#1-\pgflinewidth),
    inner sep=0pt,
    outer sep=0pt,
    ultra thick,
    rotate=45},
    plus/.default={5pt},
  cross/.style={path picture={
    \draw[black]
      (path picture bounding box.south east) --
      (path picture bounding box.north west)
      (path picture bounding box.south west) --
      (path picture bounding box.north east);
  }},
}

\pgfmathsetmacro{\xsep}{33}
\pgfmathsetmacro{\ysep}{23}
\pgfmathsetmacro{\regridy}{11}
\pgfmathsetmacro{\ysepone}{0}
\pgfmathsetmacro{\yseptwo}{2}
\pgfmathsetmacro{\ysepthree}{6}
\pgfmathsetmacro{\ysepfour}{12}
\pgfmathsetmacro{\ysepfive}{20}
\pgfmathsetmacro{\timelabelsep}{7}
\pgfmathsetmacro{\gridlabelsep}{3}

\begin{figure}
  \centering
\begin{tikzpicture}[every node/.style={inner sep=0pt}, on grid]
  \node[griddot] (dot00) {};
  \node[griddot, below=\ysep pt of dot00] (dot01) {};
  \node[griddot, below=\ysep pt of dot01] (dot02) {};
  \node[griddot, below right=\ysepone and \xsep pt of dot00] (dot10) {};
  \node[griddot, below=\ysep pt of dot10] (dot11) {};
  \node[griddot, below=\ysep pt of dot11] (dot12) {};
  \node[griddot, below right=\yseptwo pt and \xsep pt of dot10] (dot20) {};
  \node[griddot, below=\ysep pt of dot20] (dot21) {};
  \node[griddot, below=\ysep pt of dot21] (dot22) {};
  \node[griddot, below right=\ysepthree pt and \xsep pt of dot20] (dot30) {};
  \node[griddot, below=\ysep pt of dot30] (dot31) {};
  \node[griddot, below=\ysep pt of dot31] (dot32) {};
  \node[griddot, below right=\ysepfour pt and \xsep pt of dot30] (dot40) {};
  \node[griddot, below=\ysep pt of dot40] (dot41) {};
  \node[griddot, below=\ysep pt of dot41] (dot42) {};
  \node[griddot, below right=\ysepfive pt and \xsep pt of dot40] (dot50) {};
  \node[griddot, below=\ysep pt of dot50] (dot51) {};
  \node[griddot, below=\ysep pt of dot51] (dot52) {};
  \node[interpdot, below=8pt of dot00] (gridinterp0) {};
  \node[interpdot, below=8pt of dot10] (gridinterp1) {};
  \node[interpdot, below=8pt of dot20] (gridinterp2) {};
  \node[interpdot, below=8pt of dot30] (gridinterp3) {};
  \node[interpdot, below=8pt of dot40] (gridinterp4) {};
  \node[interpdot, below=8pt of dot50] (gridinterp5) {};
  \draw[myorange]  (gridinterp0) to [out=0, in=146] (gridinterp5);
  \node[plus, mypurple, below=14pt of dot00] (interpdot0) {};
  \node[plus, mypurple, right=\xsep pt of interpdot0] (interpdot1) {};
  \node[plus, mypurple, right=\xsep pt of interpdot1] (interpdot2) {};
  \node[plus, mypurple, right=\xsep pt of interpdot2] (interpdot3) {};
  \node[plus, mypurple, right=\xsep pt of interpdot3] (interpdot4) {};
  \node[plus, mypurple, right=\xsep pt of interpdot4] (interpdot5) {};
  \draw[mypurple]  (interpdot0) -- (interpdot5);
  \node[circle, cross, right=24pt of interpdot2, minimum width=12pt] (X) {};
  \node[fill=none, above left=10pt and 10pt of dot00] (excision0) {};
  \node[fill=none, above right=10pt and 10pt of dot50] (excision1) {};
  \draw[thick, postaction={decorate, decoration={raise=5pt, text along path,
    text align=center, text={domain boundary}}}, dotted]
    (excision0) to [out=-1, in=140] (excision1);
  \node[fill=none, below right=10pt and 15pt of dot52] (xline) {};
  \node[fill=none, left=200pt of xline] (origin) {};
  \node[fill=none, above=120pt of origin] (yline) {};
  \draw[->] (origin) -- (yline) node [midway, sloped, above=5pt] {$x_{\rm{I}}$};
  \draw[->] (origin) -- (xline) node [midway, below=5pt] {$t$};
\end{tikzpicture}
  \caption[Spacetime interpolation in the inertial frame]
  {Spacetime interpolation to the black \mbox{\large $\times$},
    as viewed in the inertial frame of \SpEC{}.
    The green circles represent the grid points of the BBH simulation at the
    times
    where metric data was stored to disk.
    The dotted line corresponds to the domain boundary of the simulation.
    If we first perform a set of spatial interpolations, then interpolate
    the results in time, we have two choices for how to handle these
    interpolations.
    One choice is to interpolate to a constant location in the
    grid frame shown in orange, or a constant location in the inertial frame
    shown in purple.
    The grid frame interpolation is advantageous for multiple reasons.
  }
  \label{figInertialInterp}
\end{figure}
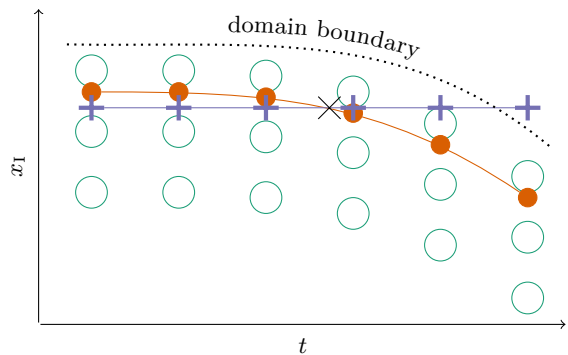

In the inertial frame, the grid points on which the scalar $A$ is defined
are moving
with respect to time, as seen in \cref{figInertialInterp}.
As a consequence, the domain boundary that is stationary in the grid
frame is also moving with time.
Suppose we are interested in the value of $A$ at the
\mbox{\large $\times$}, located at ($x^i_{\rm{I}}, t$) where $x^i_{\rm{I}}$
is the spatial location in the inertial frame,
and we want to use $6$ time slices to perform a $5$\ts{th} order time
interpolation.
If we choose to perform a spatial interpolation on each of the $6$ time slices
first, then perform a temporal interpolation to the time $t$, then we have
two choices for how to spatially interpolate.

The first choice, shown with pluses and a line in purple,
is to spatially interpolate to $x^i_{\rm{I}}$ on
each time slice, then interpolate in time.
This method has two major drawbacks.
The scalar $A$ is, by construction,
usually varying slower in time when viewed at a
constant grid point $x^i_{\rm{G}}$ compared to a constant inertial point
$x^i_{\rm{I}}$.
The result is less accurate temporal interpolations along $x^i_{\rm{I}}$ which
leads to decreased time step sizes.
In addition, spatially interpolating to a constant point in the inertial frame
could lead to attempting
to spatially interpolate outside of the domain, as seen on the last
time slice on the right side of the figure.
Therefore, the preferred option is to interpolate to a constant grid frame
point on
each time slice, then interpolate in time, as shown with filled dots and a
line in orange.

\begin{figure}
  \centering
\begin{tikzpicture}[every node/.style={inner sep=0pt}, on grid]
  \node[griddot] (dot00) {};
  \node[griddot, below=\ysep pt of dot00] (dot01) {};
  \node[griddot, below=\ysep pt of dot01] (dot02) {};
  \node[griddot, right=\xsep pt of dot00] (dot10) {};
  \node[griddot, below=\ysep pt of dot10] (dot11) {};
  \node[griddot, below=\ysep pt of dot11] (dot12) {};
  \node[griddot, right=\xsep pt of dot10] (dot20) {};
  \node[griddot, below=\ysep pt of dot20] (dot21) {};
  \node[griddot, below=\ysep pt of dot21] (dot22) {};
  \node[griddot, right=\xsep pt of dot20] (dot30) {};
  \node[griddot, below=\ysep pt of dot30] (dot31) {};
  \node[griddot, below=\ysep pt of dot31] (dot32) {};
  \node[griddot, right=\xsep pt of dot30] (dot40) {};
  \node[griddot, below=\ysep pt of dot40] (dot41) {};
  \node[griddot, below=\ysep pt of dot41] (dot42) {};
  \node[griddot, right=\xsep pt of dot40] (dot50) {};
  \node[griddot, below=\ysep pt of dot50] (dot51) {};
  \node[griddot, below=\ysep pt of dot51] (dot52) {};
  \node[interpdot, below=14pt of dot00] (interpdot0) {};
  \node[interpdot, right=\xsep pt of interpdot0] (interpdot1) {};
  \node[interpdot, right=\xsep pt of interpdot1] (interpdot2) {};
  \node[interpdot, right=\xsep pt of interpdot2] (interpdot3) {};
  \node[interpdot, right=\xsep pt of interpdot3] (interpdot4) {};
  \node[interpdot, right=\xsep pt of interpdot4] (interpdot5) {};
  \pgfmathsetmacro{\interpx}{24}
  \node[circle, cross, right=\interpx pt of interpdot2, minimum width=12pt] (X) {};
  \node[plus, mypink, right=\interpx pt of dot20] (timeinterp0) {};
  \node[plus, mypink, right=\interpx pt of dot21] (timeinterp1) {};
  \node[plus, mypink, right=\interpx pt of dot22] (timeinterp2) {};
  \draw[myorange]  (interpdot0) -- (interpdot5);
  \draw[mypink]  (timeinterp0) -- (timeinterp2);
  \node[fill=none, above left=10pt and 10pt of dot00] (excision0) {};
  \node[fill=none, above right=10pt and 10pt of dot50] (excision1) {};
  \draw[dotted, thick] (excision0) -- (excision1)
  node [midway, above=5pt] {domain boundary};
  \pgfmathsetmacro{\leftline}{20}
  \pgfmathsetmacro{\bottomline}{10}
  \node[fill=none, below left=\bottomline pt and \leftline pt of dot02] (origin) {};
  \node[fill=none, above left=20pt and \leftline pt of dot00] (yline1) {};
  \draw[->] (origin) -- (yline1) node [midway, sloped, above=5pt] {$x_{\rm{G}}$};
  \node[fill=none, below right=\bottomline pt and 15pt of dot52] (xline1) {};
  \draw[->] (origin) -- (xline1) node [midway, below=5pt] {$t$};
\end{tikzpicture}
  \caption[Spacetime interpolation in the grid frame]
  {Spacetime interpolation to the black \mbox{\large $\times$},
    as viewed in the
    grid frame of
    \SpEC{}.
    The setup is similar to \cref{figInertialInterp}, but we are observing in
    the grid frame.
    We demonstrate the additional choice between performing the spatial
    interpolations before or after temporal interpolations.
    Spatial before temporal is shown in orange, and temporal before spatial
    is shown in pink.
  }
  \label{figGridInterp}
\end{figure}
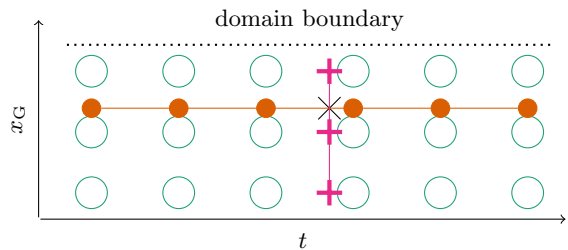

It is instructive to view this interpolation in the grid frame, as
seen in \cref{figGridInterp}.
In this frame, the locations of the domain boundary and the grid points
are stationary in time.
On each time slice, we perform a spatial interpolation to the orange points
at $x^i_{\rm{G}} = M(x^i_{\rm{I}}, t)$, where $t$ is the time to which we are
interpolating and $M$ is the time dependent mapping from the inertial
frame to the grid frame.
In this figure, we show another possibility where we first
interpolate in time along each
grid point in the subdomain to the pink pluses, then perform a spatial
interpolation.
If we count the number of operations required for either method, we find
that interpolating in time then space takes
$\mathcal{O}(N_{\rm{s}} N_t^2 + N_s^2)$
operations, where $N_{\rm{s}}$ is the number of spatial points in the
subdomain and
$N_t$ is the number of time slices used in the interpolation.
Interpolating in space then time takes $\mathcal{O}(N_t N_{\rm{s}}^2 + N_t^2)$
operations, which is typically larger than the number of operations when
interpolating in time first, since $N_{\rm{s}} > N_t$ for our case.

Unfortunately, while interpolating in time before space requires
fewer operations, in practice the error in the
interpolated tensors is larger, resulting in the generator timestepper
taking smaller steps.
We therefore default to always performing a spectral spatial interpolation on
each time slice to the grid point $x^i_{\rm{G}}$, then interpolating in time
with Lagrange polynomial interpolation.

\begin{figure}
  \centering
\begin{tikzpicture}[every node/.style={inner sep=0pt}, on grid]
  \node[griddot] (dot00) {};
  \node[griddot, below=\ysep pt of dot00] (dot01) {};
  \node[griddot, below=\ysep pt of dot01] (dot02) {};
  \node[griddot, right=\xsep pt of dot00] (dot10) {};
  \node[griddot, below=\ysep pt of dot10] (dot11) {};
  \node[griddot, below=\ysep pt of dot11] (dot12) {};
  \node[fill=none, right=\xsep pt of dot10] (dot20) {};
  \node[griddot, below=\regridy pt of dot20] (dot21) {};
  \node[griddot, below=\ysep pt of dot21] (dot22) {};
  \node[fill=none, right=\xsep pt of dot20] (dot30) {};
  \node[griddot, below=\regridy pt of dot30] (dot31) {};
  \node[griddot, below=\ysep pt of dot31] (dot32) {};
  \node[fill=none, right=\xsep pt of dot30] (dot40) {};
  \node[griddot, below=\regridy pt of dot40] (dot41) {};
  \node[griddot, below=\ysep pt of dot41] (dot42) {};
  \node[griddot, right=\xsep pt of dot40] (dot50) {};
  \node[griddot, below=\ysep pt of dot50] (dot51) {};
  \node[griddot, below=\ysep pt of dot51] (dot52) {};

  \node[above right=17pt and 12pt of dot10, fill=none] (gridChange00) {};
  \node[below right=15pt and 12pt of dot12, fill=none] (gridChange01) {};
  \draw[mypurple, dashed, very thick] (gridChange00) -- (gridChange01)
    node [pos=0.07, right=\gridlabelsep pt, black] (G2Label) {G2}
    node [below=2 pt, mypurple] (T1half) {$t_{1.5}$};

  \node[above left=17pt and 12pt of dot50, fill=none] (gridChange10) {};
  \node[below left=15pt and 12pt of dot52, fill=none] (gridChange11) {};
  \draw[mypurple, dashed, very thick] (gridChange10) -- (gridChange11)
    node [pos=0.07, right=\gridlabelsep pt, black] (G3Label) {G3}
    node [below=2 pt, mypurple] (T2half) {$t_{2.5}$};

  \node[below=7pt of dot00, interpdot] (interpdot0) {};
  \node[right=\xsep pt of interpdot0, interpdot] (interpdot1) {};
  \node[below right=17pt and \xsep pt of interpdot1, interpdot] (interpdot2) {};
  \node[right=\xsep pt of interpdot2, interpdot] (interpdot3) {};
  \node[right=\xsep pt of interpdot3, interpdot] (interpdot4) {};
  \node[above right=8pt and \xsep pt of interpdot4, interpdot] (interpdot5) {};

  \node[right=12pt of interpdot1, gridchangedot] (gridchangedot00) {};
  \node[left={\xsep-12 pt} of interpdot2, gridchangedot] (gridchangedot01) {};
  \draw[myorange] (interpdot0) to (gridchangedot00) {};

  \node[right={\xsep - 12 pt} of interpdot4, gridchangedot] (gridchangedot10) {};
  \draw[myorange] (gridchangedot01) to (gridchangedot10);
  \node[left=12pt of interpdot5, gridchangedot] (gridchangedot11) {};

  \draw[myorange] (gridchangedot11) to (interpdot5) {};

  \node[right=18pt of interpdot2, circle, cross, minimum width=12pt] (X) {};

  \node[fill=none, above left=17pt and 10pt of dot00] (excision0) {};
  \node[fill=none, above right=17pt and 10pt of dot50] (excision1) {};
  \draw[dotted, thick] (excision0) -- (excision1)
    node [midway, above=5pt] {domain boundary};

  \node[fill=none, below left=12pt and 15pt of dot02] (origin) {};
  \node[fill=none, above left=30pt and 15pt of dot00] (yline1) {};
  \draw[->] (origin) -- (yline1) node [midway, sloped, above=5pt] {$x_{\rm{G}}$}
    node [pos=0.79, right=3pt, black] (G1Label) {G1};
  \node[fill=none, below right=12pt and 15pt of dot52] (xline1) {};
  \draw[->] (origin) -- (xline1) node [midway, below=5pt] {$t$};
\end{tikzpicture}
  \caption[Spacetime interpolation in the grid frame with AMR regrids]
  {Spacetime interpolation to the black \mbox{\large $\times$},
    as viewed in the
    grid frame of
    \SpEC{}, showing spatial before temporal interpolation and constant
    grid location interpolation.
    The setup is similar to \cref{figGridInterp}, but we now have AMR.
    The vertical dashed purple lines correspond to AMR regrids, where the
    grid in general is quite different before and after the regrid.
    When we encounter a regrid, we must find the relationship between the
    regrids at the black dot locations
    by using the inertial frame which is continuous across regrids, as
    seen in \cref{figInertialInterpAMR}.
  }
  \label{figGridInterpAMR}
\end{figure}
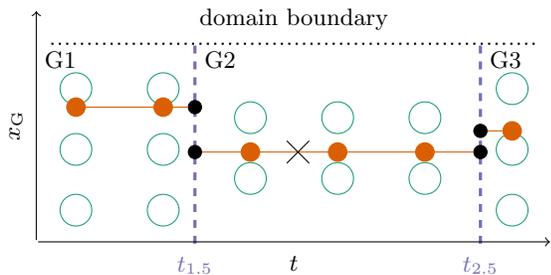

The situation becomes more complicated when adaptive mesh refinement (AMR)
during the original BBH evolution alters the grid frame.
In general, the evolution grid has a different number of points after an AMR
regrid, and the coordinates
in the grid frame are not continuous across the regrid.
In \cref{figGridInterpAMR}, we see two AMR regrids denoted by vertical
dashed lines at times $t_{1.5}$ and $t_{2.5}$.
We start in the grid frame labeled G$2$, where the desired interpolation
location
$(x^i_{\rm{I}}, t)$ lives,
following the same procedure of mapping to the grid frame location
$x^i_{\rm{G}2} = M_2(x^i_{\rm{I}}, t)$, where
$M_2$ is the mapping from the inertial frame to the grid frame G$2$.
We spatially interpolate to the grid point $x^i_{\rm{G}2}$ at all the
times within the time interpolation stencil and in the frame G$2$.
When a regrid occurs, we must determine how the two neighboring grid frames
are related so we know to what grid location to interpolate.
Specifically, we need to know what the corresponding grid frame locations
in G$1$ and G$3$ are, that is, $x^i_{\rm{G}1}$ and $x^i_{\rm{G}3}$ respectively.

We make use of the inertial frame whose coordinates are continuous across
the regrid to find the relationship between the grid frames.
Consider the regrid at $t_{1.5}$.
We map from the G$2$ grid frame location to the inertial frame via
the G$2$ mapping
$M_2^{-1}(x^i_{\rm{G}2}, t_{1.5})$, then map from the inertial frame to
the G$1$ grid to find the corresponding grid location $x^i_{\rm{G}1}$.
Therefore, the relationship between the grid locations is
\begin{equation}
  x^i_{\rm{G}1} = M_1(M^{-1}_2(x^i_{\rm{G}2}, t_{1.5}), t_{1.5}).
  \label{eqnGridToGrid}
\end{equation}
This procedure is applied at every regrid in the range of times where
temporal interpolation occurs.
The result is a set of straight lines in the grid frame shown in
\cref{figGridInterpAMR} along which we interpolate in time.

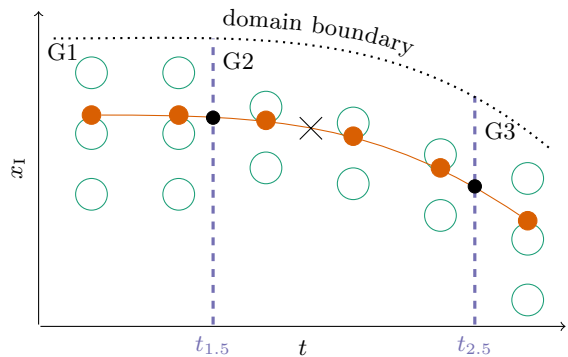
\begin{figure}
  \centering
\begin{tikzpicture}[every node/.style={inner sep=0pt}, on grid]
  \node[griddot] (dot00) {};
  \node[griddot, below=\ysep pt of dot00] (dot01) {};
  \node[griddot, below=\ysep pt of dot01] (dot02) {};
  \node[griddot, below right=\ysepone and \xsep pt of dot00] (dot10) {};
  \node[griddot, below=\ysep pt of dot10] (dot11) {};
  \node[griddot, below=\ysep pt of dot11] (dot12) {};
  \node[fill=none, below right=\yseptwo pt and \xsep pt of dot10] (dot20) {};
  \node[griddot, below=\regridy pt of dot20] (dot21) {};
  \node[griddot, below=\ysep pt of dot21] (dot22) {};
  \node[fill=none, below right=\ysepthree pt and \xsep pt of dot20] (dot30) {};
  \node[griddot, below=\regridy pt of dot30] (dot31) {};
  \node[griddot, below=\ysep pt of dot31] (dot32) {};
  \node[fill=none, below right=\ysepfour pt and \xsep pt of dot30] (dot40) {};
  \node[griddot, below=\regridy pt of dot40] (dot41) {};
  \node[griddot, below=\ysep pt of dot41] (dot42) {};
  \node[griddot, below right=\ysepfive pt and \xsep pt of dot40] (dot50) {};
  \node[griddot, below=\ysep pt of dot50] (dot51) {};
  \node[griddot, below=\ysep pt of dot51] (dot52) {};

  \node[fill=none, below right=10pt and 15pt of dot52] (xline) {};
  \node[fill=none, left=200pt of xline] (origin) {};
  \node[fill=none, above=120pt of origin] (yline) {};
  \draw[->] (origin) -- (yline) node [midway, sloped, above=5pt] {$x_{\rm{I}}$}
    node [pos=0.87, right=3pt, black] (G1Label) {G1};
  \draw[->] (origin) -- (xline) node [midway, below=5pt] {$t$};

  \node[right={2.*\xsep pt} of origin, fill=none] (gridChange00) {};
  \node[above={4.77*\ysep pt} of gridChange00, fill=none] (gridChange01) {};
  \draw[mypurple, dashed, very thick] (gridChange00) -- (gridChange01)
    node [pos=0.92, right=\gridlabelsep pt, black] (G2Label) {G2}
    node [below=\timelabelsep pt of gridChange00, mypurple] (T1half) {$t_{1.5}$};

  \node[right={3.*\xsep pt} of gridChange00, fill=none] (gridChange10) {};
  \node[above={3.82*\ysep pt} of gridChange10, fill=none] (gridChange11) {};
  \draw[mypurple, dashed, very thick] (gridChange10) -- (gridChange11)
    node [pos=0.85, right=\gridlabelsep pt, black] (G3Label) {G3}
    node [below=\timelabelsep pt of gridChange10, mypurple] (T2half) {$t_{2.5}$};

  \pgfmathsetmacro{\yinterp}{16}
  \node[interpdot, below=\yinterp pt of dot00] (gridinterp0) {};
  \node[interpdot, below=\yinterp pt of dot10] (gridinterp1) {};
  \node[interpdot, below=\yinterp pt of dot20] (gridinterp2) {};
  \node[interpdot, below=\yinterp pt of dot30] (gridinterp3) {};
  \node[interpdot, below=\yinterp pt of dot40] (gridinterp4) {};
  \node[interpdot, below=\yinterp pt of dot50] (gridinterp5) {};
  \draw[myorange]  (gridinterp0) to [out=0, in=146] (gridinterp5);

  \node[below right=8pt and 17pt of dot21, circle, cross, minimum width=12pt] (X) {};

  \node[below right=1pt and 13pt of gridinterp1, gridchangedot] (gridchangedot00) {};
  \node[above left=13pt and 20pt of gridinterp5, gridchangedot] (gridchangedot11) {};

  \node[fill=none, above left=13pt and 15pt of dot00] (excision0) {};
  \node[fill=none, above right=10pt and 10pt of dot50] (excision1) {};
  \draw[thick, postaction={decorate, decoration={raise=5pt, text along path, text align=center, text={domain boundary}}}, dotted]  (excision0) to [out=0, in=140]
    (excision1);
\end{tikzpicture}
  \caption[Spacetime interpolation in the inertial frame with AMR regrids]
  {Spacetime interpolation to the black \mbox{\large $\times$}
    as in \cref{figGridInterpAMR}, but viewed in the inertial frame of
    \SpEC{}.
    The line along which we are interpolating is continuous in the inertial
    frame, and the black dots on the boundary between regrids are used to
    find how the neighboring grid frames are related.
  }
  \label{figInertialInterpAMR}
\end{figure}

The corresponding inertial frame viewpoint is shown in
\cref{figInertialInterpAMR}.
Again we see that the domain boundary and grid points are in general at
different locations in the inertial frame, but the line along which we
are interpolating
is continuous across the regrids unlike in the grid frame.
The black dot at each regrid time is used as the anchor point to map between
the neighboring grid frames in \cref{eqnGridToGrid}.
Specifically, the black dot along the first regrid
satisfies
\begin{equation}
  M^{-1}_1(x^i_{\rm{G}1}, t_{1.5}) = M^{-1}_2(x^i_{\rm{G}2}, t_{1.5}).
  \label{eqnMapBetweenGrids}
\end{equation}

\begin{figure}
  \centering
  \includegraphics[width=0.9\columnwidth]{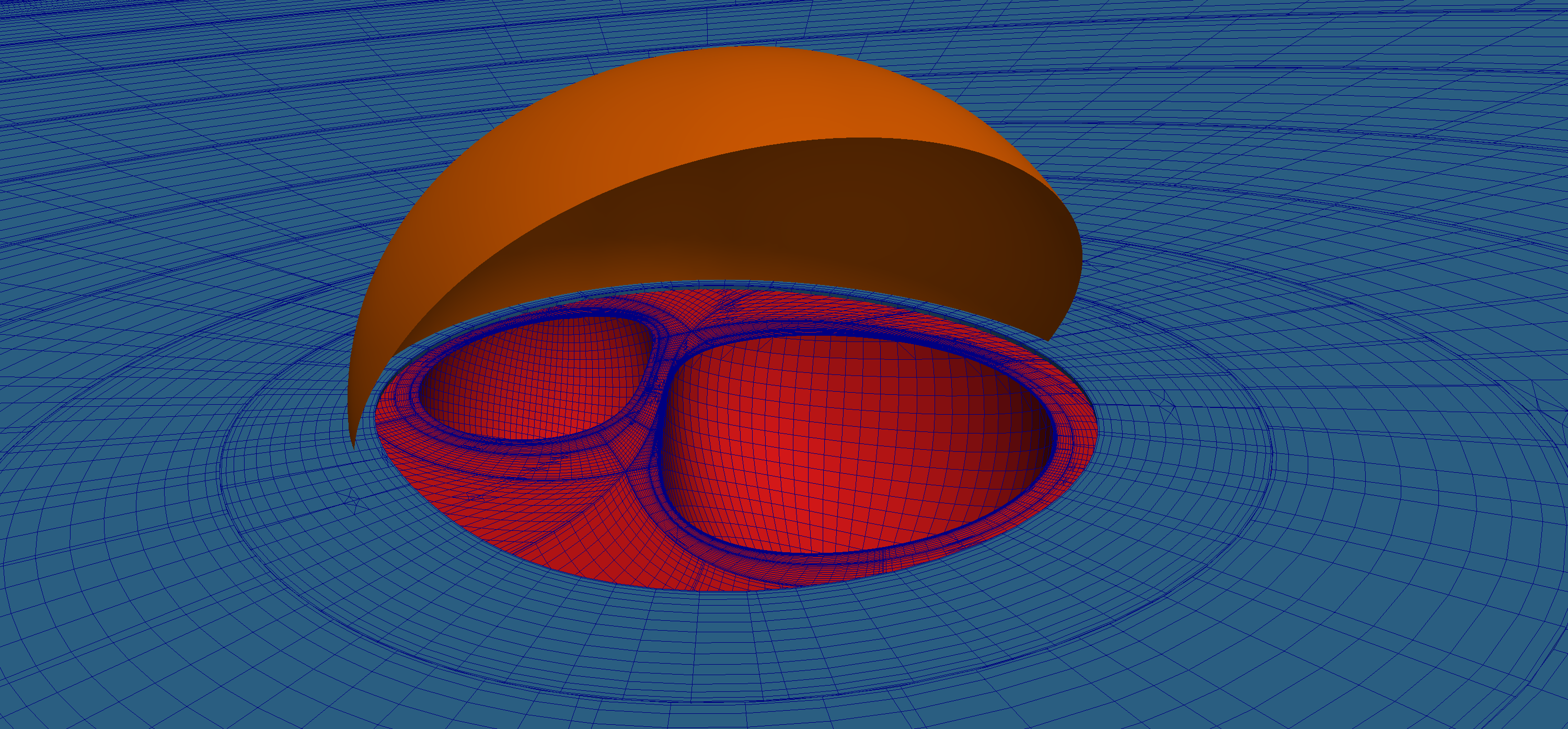}
  \caption[BBH Evolution domain with apparent horizons excised and event horizon
    shown]
  {Portion of the event horizon surface, shown in orange, on top of the
    \SpEC{} domain structure just before and after the grid change for the
    ringdown.
    In red, we see the evolution grid just before the ringdown grid change
    including the excision regions associated with the two inner
    apparent horizons.
    In blue, we see the evolution grid just after the ringdown grid change,
    with only one excision region associated with the common apparent horizon.
    Both the inner and common apparent horizons can be seen in panel~(d)
    of \cref{figThorneSuggestion}.
  }
  \label{figGW150914}
\end{figure}

There is an additional complication to this procedure, albeit rare, that
can occur when determining the relationship between neighboring grid
frames in \cref{eqnGridToGrid}.
In \cref{figGW150914}, we see part of the domain structure for
a BBH simulation with parameters consistent with
the Advanced LIGO event~\cite{Abbott:2016blz}, specifically
$m_1/m_2 = 1.25$ with dimensionless spin magnitudes
$\chi_1 = 0.45$, $\chi_2 = 0.54$ in arbitrary
directions.
In red, we see a cutaway of the inspiral domain structure just before the domain
topology changes for the ringdown, where there are two excision regions
associated with individual apparent horizons of the black holes.
At this time, \SpEC{} finds a common apparent horizon encapsulating both of the
inner apparent horizons, which triggers the evolution domain to change
topology to
have just one excision region.
The new domain structure after the regrid is shown in blue, so all the
structure near the inner apparent horizons shown
in red has been excised from the domain.
Finally, in orange, we show a portion of the event horizon surface.
Since the apparent horizon is never outside the event horizon, and the excision
region by construction is always inside the apparent horizon,
the event horizon surface always encapsulates the excision region completely.

Consider the transition between $\rm{G}2$ to $\rm{G}3$ in
\cref{figInertialInterpAMR},
and assume that this transition is associated with the domain change from
the red inspiral grid to the blue ringdown grid.
If the point to which we want to interpolate resides in the
red region after the regrid, then the point will be off the domain,
causing the interpolation to fail.
In \SpEC{}, regrids can only cause grid locations to be removed from the
evolution
grid, not to enter the evolution grid.
Therefore, we use a lopsided time interpolation stencil favoring earlier
time slices to solve this issue.
We first try a balanced
stencil with $n/2$ times on either side of the desired interpolation
point,
and retry with $n/2+1$ times before the point and $n/2-1$ after if it
fails, and so forth.

While we have the ability to perform spacetime interpolations
in multiple ways, the default is to interpolate first in space to a constant
point in the grid frame on each time slice required for the time interpolation,
then perform the time interpolation.
The primary advantages to these choices are that the code handles
interpolation requests accurately and without failure
near domain boundaries and AMR regrids.

\section{Removing triangles from the collision detection algorithm}
\label{appendixCollisionDetection}

When tracing event horizon generators backwards through time, generators
leave the EH surface when they meet other generators.
These meeting points are classified as either caustics where neighboring
generators meet or crossover points where non-neighboring generators meet.
We detect crossover points by searching for EH surface self-intersections
where in theory two generators cross, but in practice we only identify
that a generator $q$ intersected a triangle $\triangle abc$ between neighboring
times $t_0$ and $t_1$ as described in \cref{secIdentifyingGenerators}.
This collision implies there is some EH generator $u$ (that we
have not evolved) inside $\triangle abc$
that met with $q$ between $t_0$ and $t_1$, so we flag $q$ as
leaving the horizon backwards through time.

\begin{figure}
  \centering
  \includegraphics[width=0.5\textwidth]{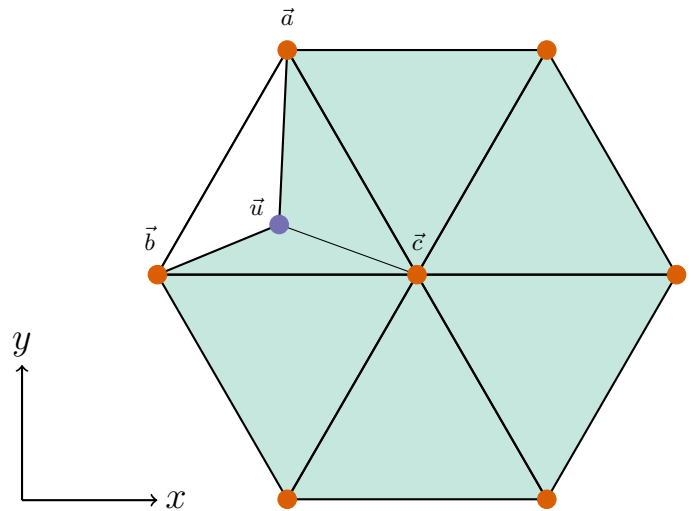}
  \caption[Intersecting plane waves toy model to study the collision detection
    algorithm]
  {A portion of a null plane wave, with normal out of the page,
    approximated by a set of null generators shown as orange dots.
    This null plane wave and another null plane wave, not shown,
    connect to form a toy event horizon used to study the collision
    detection algorithm.
    The shaded blue-green triangles are filled with future generators of this
    toy event horizon, where a hypothetical generator of the EH shown
    as a purple dot, $u$,
    has just converted from a future generator to a true generator at this
    time.
  }
  \label{figCDAddNew}
\end{figure}

Consider the setup in \cref{figCDAddNew} where we follow part of
a null plane wave
satisfying $t = z$
approximated by a set of generators in orange dots connected
to form a set of triangles.
Another null plane wave, not shown, satisfies $t = (x-y-z)/\sqrt{3}$
and is similarly approximated by a set of generators.
On small scales, these two intersecting plane waves
roughly approximate two
intersecting portions of the EH surface.
We want to search for intersections of these plane waves using only the
generators of the plane waves we are evolving.
We know analytically that the intersection of these waves in the
plane of \cref{figCDAddNew} satisfies $t = (x-y)/(1+\sqrt{3})$
and so travels in the $(+x, -y)$ direction forwards
in time (and travels faster than the speed of light).
After the two waves intersect, the future generators shaded with blue-green
will join the event horizon.
At this particular time, a generator of the plane wave not shown in this
figure, $q$, intersects $\triangle abc$ at the location $u$, so a generator
at $u$ would join the horizon at this instant along with $q$.

As was done in \cref{secIdentifyingGenerators}, the algorithm is to
follow both plane waves backwards through time to search for intersections
where generators leave the surface.
We need to identify for each generator we keep track of, shown as
an orange dot,
when the generator leaves the horizon.
One way to handle the fact that the generator $q$
intersected a hypothetical generator at $u$ is to actually create a
new generator at $u$ and keep track of it.
As shown in \cref{figCDAddNew}, we would then classify
$\triangle acu$ and $\triangle ucb$ as being filled with future generators,
and $\triangle aub$ would still be part of the EH surface.
We should therefore remove $\triangle acu$ and $\triangle ucb$ from the
collision detection algorithm, and leave $\triangle aub$ in the algorithm,
because we only want to detect collisions between generators that are
both on the EH.

This method would give a correct algorithm, but introduces some
additional complications, so we seek a simpler algorithm.
Without adding a generator, is it better to
continue to include $\triangle abc$ in the
algorithm or remove it from the algorithm?
Both choices have some potential failure modes we need to consider.
If we continue to include the triangle in the collision detection, then
the potential failure mode occurs when some generator $w$ intersects
either $\triangle acu$ or $\triangle ucb$, and we proceed to incorrectly
flag $w$ as having left the EH backwards in time.
The generator $w$ should still be considered part of the EH surface
because we only care about surface self-intersections between
two generators that are both on the EH.
However, in this setup of two colliding plane waves, there will never be
such a generator $w$ that is falsely flagged, because the plane wave
to which $w$ belongs has already passed by the triangles $\triangle acu$
and $\triangle ucb$.
Therefore, including the full triangle $\triangle abc$ in the
collision detection algorithm introduces no failure modes that are
possible if the EH is sufficiently covered with generators.

The other option is to remove $\triangle abc$ from the algorithm.
The potential failure mode here occurs when a generator $w$
should have intersected some generator in $\triangle aub$ causing
it to leave the horizon,
but we incorrectly label $w$ as still being a part of the EH surface.
This failure mode can and does occur in both this toy model example and
in realistic BBH event horizon simulations.
Therefore, removing the triangle from the collision detection yields
incorrect results, where some generators are falsely flagged as
being on the EH.

To summarize, the method we use is to keep $\triangle abc$ in the
collision detection algorithm until the entire triangle is filled
with future generators, or equivalently when all three generators
$a$, $b$, and $c$ are all flagged as future generators of the EH.
If all three generators that form the triangle are future generators,
the triangle must be removed from the algorithm.
This is because the approximation of two intersecting plane waves breaks
down on large or long timescales, so it is possible for the triangle
of future generators to wrap back toward the EH as we trace it backwards
through time.
If the triangle is never removed from the algorithm, then we see
some future generator triangles intersecting with generators on the EH
surface, resulting in unphysical holes in the event horizon.

\begin{acknowledgments}
We thank David Nichols for useful conversations about affine parametrizations
of event horizon generators.
We are grateful to Fran\c{c}ois H\'{e}bert and William Throwe for various
helpful conversations including numerous triangle drawing whiteboard sessions.
We thank Daniel A. Hemberger for helping us understand the intricacies
of the adaptive mesh refinement in \SpEC{}.
For helping smooth the visualization
of event horizon surfaces, we thank Curran D. Muhlberger.
We also thank Harald Pfeiffer for providing the BBH simulation with parameters
similar to the system detected by Advanced LIGO, shown in \cref{figGW150914}.
For providing useful suggestions during the writing phase, we thank
Nils Deppe.

We gratefully acknowledge support for this research at Cornell from the
Sherman Fairchild Foundation and NSF grants PHY-1306125 and AST-1333129.
Calculations were performed on the Zwicky cluster at Caltech,
which is supported by the Sherman Fairchild Foundation and by
NSF award PHY-0960291; on the NFS XSEDE network under grant TG-PHY990007N;
at the GPC supercomputer at the SciNet HPC Consortium \cite{Scinet};
SciNet is funded by:
the Canada Foundation for Innovation (CFI) under the
auspices of Compute Canada; the Government of Ontario;
Ontario Research Fund (ORF) – Research Excellence;
and the University of Toronto.
All the surface visualizations were done using Paraview~\cite{paraviewweb}.

\end{acknowledgments}

\bibliography{References/References}

\end{document}